\documentclass{ws-ijmpa}
\usepackage[super]{cite}
\usepackage{graphicx}
\usepackage{amsmath}
\usepackage{amssymb}
\usepackage{mathrsfs}

\usepackage[colorinlistoftodos]{todonotes}

\begin{document}

\markboth{D. E. Krause {\em et al.}}
{Phenomenological Implications of a Magnetic 5th Force}

\catchline{}{}{}{}{}

\title{Phenomenological Implications of a Magnetic 5th Force}

\author{Dennis E. Krause}

\address{Physics Department, Wabash College, Crawfordsville, IN 47933, USA \\
	Department of Physics  and Astronomy, Purdue University, West Lafayette, IN 47907, USA}

\author{Joseph Bertaux}
\address{Department of Physics  and Astronomy, Purdue University, West Lafayette, IN 47907, USA}

\author{A. Meenakshi McNamara}
\address{Department of Physics  and Astronomy, Purdue University, West Lafayette, IN 47907, USA}

\author{John T. Gruenwald}
\address{Department of Physics  and Astronomy, Purdue University, West Lafayette, IN 47907, USA
\\
Snare, Inc., West Lafayette IN, 47907 USA}

\author{Andrew Longman}
\address{Department of Physics  and Astronomy, Purdue University, West Lafayette, IN 47907, USA}

\author{Carol Y. Scarlett}
\address{Department of Physics, Florida A\&M University, Tallahassee, FL 32307 USA \\
Argonne National Laboratory, Lemont, IL 60439 USA}

\author{Ephraim Fischbach}
\address{Department of Physics  and Astronomy, Purdue University, West Lafayette, IN 47907, USA \\
Snare, Inc., West Lafayette IN, 47907 USA \\
ephraim@purdue.edu }

\maketitle


\begin{abstract}
	A 5th force coupling to baryon number $B$ has been proposed to account for the correlations between the acceleration differences $\Delta a_{ij}$ of the samples studied in the E\"{o}tv\"{o}s experiment, and the corresponding differences in the baryon-to-mass ratios $\Delta(B/\mu)_{ij}$. To date the E\"{o}tv\"{o}s results have not been supported by modern experiments. Here we investigate the phenomenological implications of a possible magnetic analog $\vec{\mathscr{B}}_5$ of the conventional 5th force electric field, $\vec{\mathscr{E}}_5$, arising from the Earth's rotation. We demonstrate that, in the presence of couplings proportional to $\vec{\mathscr{B}}_5$, both the magnitude and direction of a possible 5th force field could be quite different from what would otherwise be expected and warrants further investigation.

\end{abstract}

\ccode{PACS Nos.: 04.90.+e, 03.50.Kk}

	\section{Introduction}
	In 1955 Lee and Yang \cite{Lee Yang} raised the question of whether there existed in nature a long range field which coupled to baryon number $B$, in a manner similar to the coupling of the electromagnetic 4-vector potential $A^\mu(\vec{r},t)$ to the electric charge $Q$. Since $B$, like $Q$, was believed to be absolutely conserved, the analogy also suggests that a source of baryons could give rise to a 4-vector potential $\mathscr{A}_5^\mu(\vec{r},t)$, analogous to  $A^\mu(\vec{r},t)$.  They noted that since the Earth is a source of baryon number, the presence of $\mathscr{A}_{5}^\mu(\vec{r},t)$ would give rise to an additional contribution (along with gravity) in tests of the weak equivalence principle (WEP). This additional contribution would explicitly depend on the baryonic compositions of the test samples in a WEP experiment, and hence limits on the strength of any coupling to $B$ could then be inferred from the absence of any evidence for violations of the WEP. Using data from the experiment by E\"{o}tv\"{o}s, Pek\'{a}r,  Fekete (EPF) \cite{EPF,EPF Book}, which was the most stringent test of the WEP at that time, Lee and Yang set a limit on the intrinsic strength $f_{5}$ relative to gravity of the interaction for a massless gauge field, obtaining
	\begin{equation}
	    \frac{f_{5}^2}{4\pi Gm_p^2} \lesssim 10^{-5},
	\end{equation}
	where $G$ is the Newtonian gravitational constant, $m_{p}$ is the mass of the proton, and we have set $\hbar = c  =1$.
    
    The 1986 reanalysis of the EPF experiment \cite{Fischbach,Fischbach AoP} expanded upon the Lee-Yang formalism by allowing the quanta of the presumed baryon number field to have a nonzero mass, which introduces a much richer landscape of possibilities for a phenomenology which can be compared to experiment.  Ref.~\citen{Fischbach} also analyzed the actual results contained in the body of the EPF paper, rather than the data appearing in a combined summary at the end of the EPF paper.  The significance of the differences between these two datasets is discussed in detail in Refs~\citen{Fischbach,Fischbach AoP,Fischbach history,Fischbach Szabo}.  One finds that the published EPF paper contains text not present in a recently discovered hand-written draft of the paper by E\"{o}tv\"{o}s  himself \cite{Fischbach Szabo}.  For example, the published EPF paper \cite{EPF,EPF Book} contains a table in Section 10 summarizing the EPF results as if each test sample was compared to Pt, even though the reference sample was Cu for most cases. The water-Pt comparison was computed by writing \cite{Fischbach AoP,Fischbach history},
	\begin{equation} 
	\begin{split}
	    (\kappa_{\rm water}-\kappa_{\rm Cu}) +  (\kappa_{\rm Cu}-\kappa_{\rm Pt}) &= (\kappa_{\rm water}-\kappa_{\rm Pt}),
	    \\
	  [-(10 \pm 2) + (4\pm 2)]\times 10^{-9} &= [-6\pm \sqrt{2^{2} + 2^{2}}]\times 10^{-9} = (-6\pm 3)\times 10^{-9},
	  \end{split}
	 \label{Pt Cu}
	\end{equation}
	where $\kappa_{i}$ is the relative difference between the gravitational and inertial masses of the $i$th sample.  It follows from Eq.~(\ref{Pt Cu}) that by referring all of the original EPF data to Pt, a potentially significant $5\sigma$  apparent violation of the WEP depending on baryon number was reduced to a less significant $2\sigma$ effect in the published result used by Lee and Yang.  Since there is no compelling physics motivation for presenting their data as described above, it is possible to assume that the summary table in Ref.~\citen{EPF} was constructed precisely to suppress any suggestion of a WEP-violating effect in the new EPF data \cite{Fischbach PoS}.

 Motivated by the implication of Ref.~\citen{Fischbach} of the existence of a new long-range ``5th force'' coupling to baryon number, a large number of experiments have been undertaken to detect the presence of such an interaction, almost all of which have given null results \cite{Fischbach_book,Fischbach Metrologia,Franklin}. The single exception, that of the floating ball experiment by Thieberger \cite{Thieberger}, will be discussed below.  However, this work led to the recognition that the search for new macroscopic-ranged forces is a valuable tool for  exploring physics beyond the Standard Model, and this continues through the present day \cite{Adelberger ISL,Adelberger Torsion,Lanfranchi 2010,Antoniadis,Wagner,Murata,Safronova,Tino,Lanfranchi 2021}.
	
Despite the significant theoretical and experimental effort in this area, it is curious that an aspect that was raised in the original Lee-Yang paper remains essentially unexplored.  In their model of a massless gauge theory coupling to baryon number, Lee and Yang mention the spin-dependent force arising from the magnetic analog to the usual magnetic field.  However, they do not pursue it further since, they note, macroscopic matter is usually unpolarized or rotating too slowly.  It is the purpose of the present paper to extend the work of Lee and Yang  and explore the phenomenological consequences of a classical magnetic field accompanying a new vector field coupling to baryon number.  However, here we will assume that the boson has mass, which breaks the explicit gauge independence of the original Lee-Yang theory.

We will begin by obtaining the general formalism of the theory, which follows closely classical electrodynamics with a massive photon.  Since we are interested in the effects of the new force on terrestrial experiments, we use this formalism to obtain the new 5th force electric and magnetic fields produced by the rotating Earth.  We then transform the fields to the Earth rotating frame, which yields the analog electric and magnetic fields that would act on a terrestrial experimental apparatus.  We observe that in the rotating frame, a portion of the inertial frame 5th force magnetic field now appears as an additional contribution to the 5th force electric field.  We then review the phenomenological consequences of the magnetic 5th force on various 5th force experiments.  We also discuss a scenario of a scalar-vector model of a 5th force where the electric portion of the 5th force is suppressed, so that the magnetic 5th force becomes dominant.  In such a scenario, the 5th force would have a position- and directional-dependence quite unlike the electric 5th force, that has heretofore been the focus of 5th force experiments.  Data from previous experiments can be re-examined to constrain this scenario, and new  experiments focusing on the magnetic effects are suggested.  The appendices include the details of our calculations and a description of the scalar-vector model.

	\section{Formalism}

    \subsection{General Results}

	We begin by assuming that the 5th force arises from a massive 4-vector potential $\mathscr{A}_{5}^{\alpha}$ coupling to baryon number $B$.  For the low-energy effective theory appropriate for our purposes, we  write the interaction Lagrangian with nucleons as
	\begin{equation}
	{\cal L}_{\rm int} = f_{5}\bar{\psi}_{n}\gamma^{\alpha}\psi_{n}\mathscr{A}_{5,\alpha},
	\end{equation}
	where $\psi_{n}$ is the nucleon field and $f_{5}$ is the nucleon coupling constant.
	In the classical limit, which is the case of interest in this paper, the theory is described  essentially by the same Lagrangian density as for classical electrodynamics with a massive photon \cite{GN 1971,Tu,GN 2010,Lechner},
	\begin{equation}
	    {\cal L}_{5} = -\frac{1}{4}\mathscr{F}_{5}^{\alpha\beta}\mathscr{F}_{5,\alpha\beta}+ \frac{1}{2}\mu_{5}\mathscr{A}_{5}^{\alpha}\mathscr{A}_{5,\alpha} + f_{5} J_{5}^{\alpha}\mathscr{A}_{5,\alpha},
	\end{equation}
	where $\mu_{5}$ is the mass of the 5th force boson (``hyperphoton'') in units where $\hbar = c =1$, $J_{5}^{\alpha}$ is the baryon number 4-current,
\begin{equation}
    J^{\mu}_{5} \equiv (\rho_{5},\vec{J}_{5}),
\end{equation}
and 
	\begin{equation}
	    \mathscr{F}_{5}^{\alpha\beta} \equiv \partial^\alpha \mathscr{A}_{5}^{\beta} - \partial^\beta \mathscr{A}_{5}^{\alpha}.
	\end{equation}
	We will assume that baryon number is conserved,
	\begin{equation}
	\partial_{\alpha}J^{\alpha}_{5} = 0,
	\label{continuity}
	 \end{equation}
	which then requires that $\mathscr{A}_{5}^{\alpha}$ satisfies the Lorenz condition as in the case of massive electrodynamics \cite{GN 2010},
	\begin{equation}
	    \partial_{\alpha}\mathscr{A}_{5}^{\alpha} = 0.
	    \label{Lorenz}
	\end{equation}
	
	Let us now define the ``hyperelectric'' and ``hypermagnetic'' fields as
	\begin{subequations}
	\begin{align}\label{fields}
		\vec{\mathscr{E}}_5 &\equiv -\vec{\nabla}\Phi_{5} - \frac{\partial\vec{\mathscr{A}}_5}{\partial t},
		\\
		\vec{\mathscr{B}}_5 &\equiv \vec{\nabla}\times\vec{\mathscr{A}}_5,
		\label{B curl A}
	\end{align}
	\end{subequations}
where $\Phi_{5} \equiv \mathscr{A}^{0}_{5}$.  These fields then satisfy the Proca field equations \cite{GN 1971,Tu,GN 2010,Lechner}:
\begin{subequations}
	\begin{align}\label{proca}
		\vec{\nabla}\cdot\vec{\mathscr{E}}_5 &= f_{5}{\rho}_{5} - \mu_5^{2}\Phi_{5},  \\
		\vec{\nabla}\cdot\vec{\mathscr{B}}_5 &= 0, \\
		    	\vec{\nabla}\times\vec{\mathscr{E}}_5 &= -\frac{\partial\vec{\mathscr{B}}_5}{\partial t},\\ \label{curl}
		    	\vec{\nabla}\times\vec{\mathscr{B}}_5 &= f_{5}\vec{J}_{5} + \frac{\partial \vec{\mathscr{E}}_5}{\partial t} - \mu_5^{2}\vec{\mathscr{A}}_5.
	\end{align}
	\end{subequations}

The force $\vec{F}$ on a single nucleon with coupling $f_{5}$ moving with velocity $\vec{v}$ is then
\begin{equation}\label{force}
    \vec{F} = f_{5}\left(\vec{\mathscr{E}}_{5} + \vec{v}\times \vec{\mathscr{B}}_{5}\right).
\end{equation}

\subsection{Stationary Sources and Fields}

In this paper, our focus will be on experiments which are sensitive to stationary sources and fields.  The time-independent field equations for the hyperelectric and hypermagnetic fields decouple, giving
\begin{subequations}
	\begin{align}
		\vec{\nabla}\cdot\vec{\mathscr{E}}_5(\vec{r}) &= f_{5}\rho_{5}(\vec{r}) - \mu_5^{2}\Phi_{5}(\vec{r}), 
		\label{static E Gauss}\\
		    	\vec{\nabla}\times\vec{\mathscr{E}}_5(\vec{r}) &=0,
	\end{align}
\end{subequations}
and
\begin{subequations}
	\begin{align}
	\vec{\nabla}\cdot\vec{\mathscr{B}}_5(\vec{r}) &=0,\\
		\vec{\nabla}\times\vec{\mathscr{B}}_5(\vec{r}) &= f_{5}\vec{J}_{5}(\vec{r}) - \mu_5^{2}\vec{\mathscr{A}}_5(\vec{r}),
	\label{static Ampere}
	\end{align}
\end{subequations}
where now
\begin{equation}
    \vec{\mathscr{E}}_5(\vec{r}) = -\vec{\nabla}\Phi_{5}(\vec{r}).
\label{static E phi}
\end{equation}
(Radiation solutions arising from time-dependent sources of massive vector fields  were investigated in Refs.~\citen{Krause,Poddar 2019, Poddar 2020,Poddar 2021}.)  For these stationary fields, the Lorenz condition, Eq.~(\ref{Lorenz}), becomes
\begin{equation}
    \vec{\nabla}\cdot\vec{\mathscr{A}}_{5}(\vec{r}) = 0,
\end{equation}
while for stationary sources, the continuity equation, Eq.~(\ref{continuity}), reduces to
\begin{equation}
     \vec{\nabla}\cdot\vec{J}_{5}(\vec{r}) = 0.
\end{equation}

Using Eq.~(\ref{static E phi}), Gauss's law for the hyperelectric field Eq.~(\ref{static E Gauss}) can be rewritten as
\begin{equation}
    \left(\vec{\nabla}^{2} - \mu_{5}^{2}\right)\Phi_{5}(\vec{r}) = -f_{5}\rho_{5}(\vec{r}).
\label{static phi eqn}
\end{equation}
Similarly, after expressing the hypermagnetic field $\vec{\mathscr{B}}_{5}(\vec{r})$ in terms of $\vec{\mathscr{A}}_{5}(\vec{r})$ using Eq.~(\ref{B curl A}), the hypermagnetic Ampere's law Eq.~(\ref{static Ampere}) becomes 
\begin{equation}
    \left(\vec{\nabla}^{2} - \mu_{5}^{2}\right)\vec{\mathscr{A}}_{5}(\vec{r}) = -f_{5}\vec{J}_{5}(\vec{r}).
\label{static A eqn}
\end{equation}

\subsection{Green's Function Solutions}	
	
The static field equations for $\Phi_{5}(\vec{r})$ and $\vec{\mathscr{A}}_{5}(\vec{r})$ given by Eqs.~(\ref{static phi eqn}) and (\ref{static A eqn})  are each of the form of the modified Helmholtz equation.  Their general solutions  can then be obtained using the method of Green's functions.  The solution to the equation
	\begin{equation}
		\left(\vec{\nabla}^{2}- \mu_5^{2}\right)G(\vec{r},\vec{r}\,') = -\delta(\vec{r}-\vec{r}\,'),
	\end{equation}
with the boundary condition that the solution vanishes at infinity, is given by
\begin{equation}\label{green_function}
		G(\vec{r}, \vec{r}\,') = 
		 \frac{e^{-\mu_5|\vec{r}-\vec{r}\,'|}}{4\pi|\vec{r}-\vec{r}\,'|}.
	\end{equation}
Therefore, the general solutions to Eqs.~(\ref{static phi eqn}) and (\ref{static A eqn}) for $\Phi_{5}(\vec{r})$ and $\vec{\mathscr{A}}_{5}(\vec{r})$ for localized source densities are
\begin{subequations}
\label{static Phi A}
\begin{align}
    \Phi_{5}(\vec{r})  &= \frac{f_{5}}{4\pi} \int d^{3}r'\,\frac{e^{-\mu_5|\vec{r}-\vec{r}\,'|}}{|\vec{r}-\vec{r}\,'|}\rho_{5}(\vec{r}\,'),
\label{static Phi}
    \\
    \vec{\mathscr{A}}_{5}(\vec{r})  &= \frac{f_{5}}{4\pi} \int d^{3}r'\,\frac{e^{-\mu_5|\vec{r}-\vec{r}\,'|}}{|\vec{r}-\vec{r}\,'|}\vec{J}_{5}(\vec{r}\,').
\label{static A}
\end{align}
\end{subequations}
For a point particle at rest at the origin, $\rho_{5}(\vec{r}\,') = \delta^{3}(\vec{r}\,')$, and we obtain, as expected, the Yukawa potential
\begin{equation}
     \Phi_{5}(\vec{r}) = \frac{f_{5}}{4\pi}\frac{e^{-\mu_{5}r}}{r},
\end{equation}
where $r = |\vec{r}|$. 

Below, where we consider a source with azimuthal symmetry, it will prove convenient to use spherical coordinates.  The expansion of $G(\vec{r},\vec{r}\,')$ in terms of spherical harmonics \cite{Arfkin}, is then given by,
	\begin{equation}
	\frac{e^{-\mu_{5}|\vec{r}-\vec{r}\,'|}}{4\pi|\vec{r}-\vec{r}\,'|} = \mu_{5}\sum_{l = 0}^{\infty}i_{l}(\mu_{5}r_{<})k_{l}(\mu_{5}r_{>})\sum_{m = -l}^{l}Y_{l}^{m*}(\theta',\phi')Y_{l}^{m}(\theta,\phi),
	\label{Ylm expansion}
	\end{equation}
where $i_{l}(x)$ and $k_{l}(x)$ are modified spherical Bessel functions of the first and second kind, respectively, and $r_{>}$ ($r_{<}$) is the larger (lesser) of $r$ and $r'$.

\section{5th Force Fields of a Rotating Earth}

\subsection{Sources}

\begin{figure}[t]
    \centering
    \includegraphics[scale=.5]{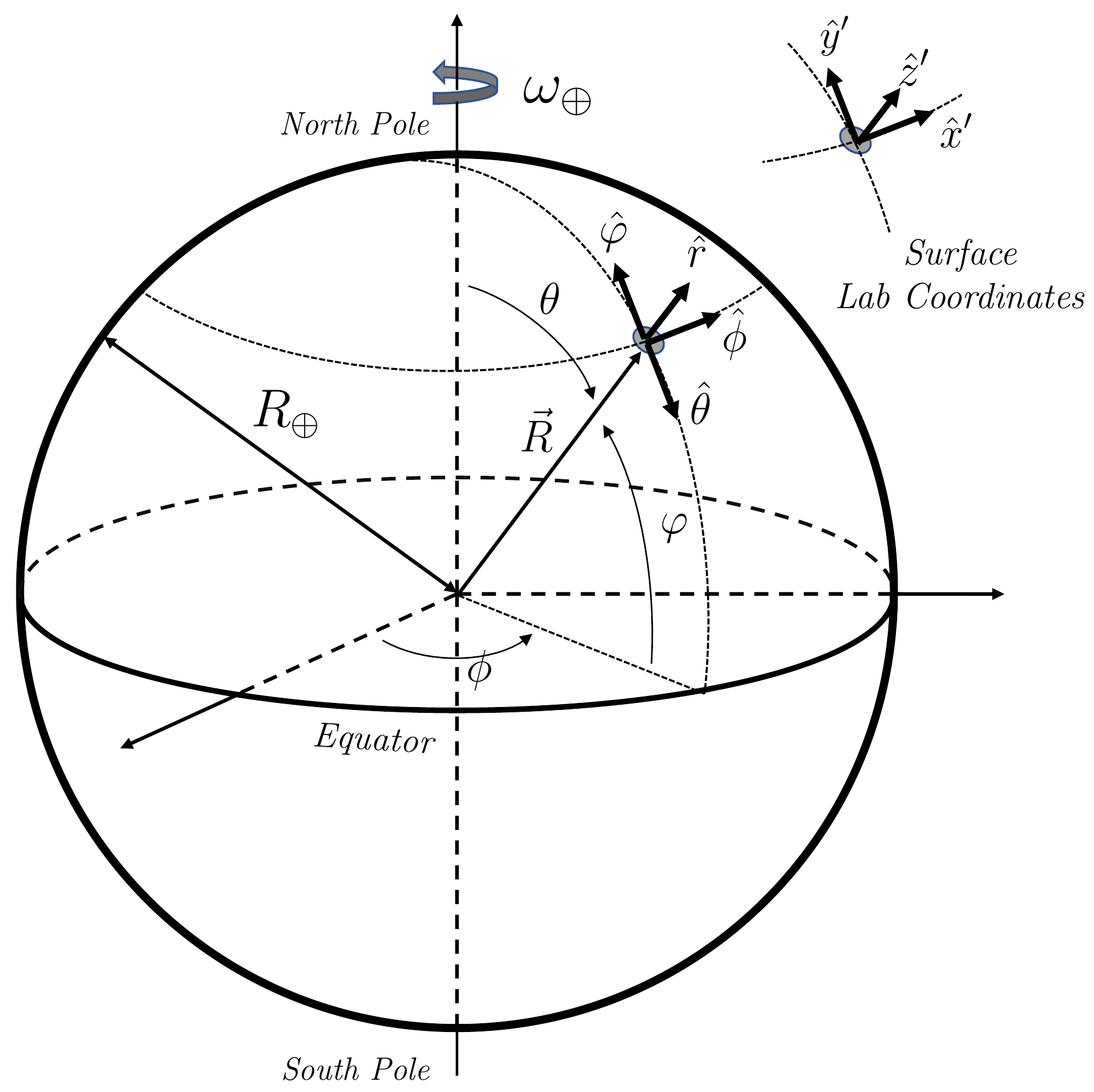}
    \caption{Idealized spherical model of the Earth showing the coordinate systems used in the text.  The shaded ovals indicate a location on the Earth's surface located at $\vec{R}$. Here $r, \theta, \phi$ denote the usual radial, polar, and azimuthal spherical coordinates, respectively. The primed unit vectors denote the axes for the surface lab frame: $\hat{x}'$ (east), $\hat{y}'$ (north), and $\hat{z}'$ (vertical).}
    \label{fig:Earth}
\end{figure}

Let us now apply the formalism developed in the previous section to the Earth as a rotating source.  For our purposes, it is sufficient to treat the Earth as a uniform ball of radius $R_{\oplus}$ rotating at constant angular speed $\omega_{\oplus}$, as shown in Fig.~\ref{fig:Earth}. 
We will assume that the baryon number density is given by 
\begin{equation}
    \rho_{5}(\vec{r}\,') = \left\{\begin{array}{ll}
        \rho_{5,\oplus} \equiv \displaystyle \frac{3B_{\oplus}}{4\pi R_{\oplus}^{3}}, & r'\leq R_{\oplus}, \\
        &\\
        0, & r' > R_{\oplus},
    \end{array}\right.
\label{Earth rho}
\end{equation}
where $B_{\oplus}$ is the total baryon number of the Earth.  The Earth's rotation creates a current density given by
\begin{equation}
    \vec{J}_{5}(\vec{r}) = \left\{\begin{array}{ll}
        \rho_{5,\oplus}\vec{\omega}_{\oplus} \times \vec{r}\,' = \rho_{5,\oplus}\omega_{\oplus}r'\sin\theta'\,\hat{\phi}' ,  & r'\leq R_{\oplus}, \\
        &\\
        0, & r' > R_{\oplus}.
    \end{array}\right.
\label{Earth J}
\end{equation}

\subsection{Potentials and Fields}

In \ref{Potentials Fields Appendix}, we calculate the potentials and fields arising from Eqs.~(\ref{Earth rho}) and (\ref{Earth J}) for $r > R_{\oplus}$ using Eq.~(\ref{static Phi A}).  For the potentials, we find
\begin{align}
\Phi_{5}(\vec{r}) &= \frac{f_{5}\rho_{5,\oplus}}{\mu_{5}^{2}}
K_{E}(\mu_{5}R_{\oplus})
\left(\frac{e^{-\mu_{5}r}}{\mu_{5}r}\right),
\label{earth Phi final} \\
\vec{\mathscr{A}}_{5}(\vec{r})  
&= f_{5}\rho_{5,\oplus}\omega_{\oplus}
K_{B}(\mu_{5}R_{\oplus})
\left[\frac{e^{-\mu_{5}r}(1 + \mu_{5}r)}{\mu_{5}^{5}r^{2}}\right]
\sin\theta\,\hat{\phi},
\label{earth A final}
\end{align}
where
\begin{equation}
     K_{E}(x) \equiv x\cosh x - \sinh x \simeq
     \left\{\begin{array}{ll}
         \displaystyle \frac{x^{3}}{3} + \frac{x^{5}}{30},  & \mbox{if $x \ll 1$}, \\
         & \\
         \displaystyle \frac{1}{2}xe^{x}, & \mbox{if $x \gg 1$},
     \end{array} 
     \right.
\end{equation}
and
\begin{equation}
     K_{B}(x) \equiv (3 + x^{2})\sinh x - 3x\cosh x \simeq
     \left\{\begin{array}{ll}
         \displaystyle \frac{x^{5}}{15} + \frac{x^{7}}{210},  & \mbox{if $x \ll 1$}, \\
         & \\
         \displaystyle \frac{1}{2}x^{2}e^{x}, & \mbox{if $x \gg 1$}.
     \end{array} 
     \right.
\end{equation}
The hyperelectric and hypermagnetic fields for $r > R_{\oplus}$ are then obtained using Eqs.~(\ref{static E phi}) and (\ref{B curl A}), respectively,
\begin{align}
\vec{\mathscr{E}}_5(\vec{r}) &=\frac{f_{5}\rho_{5,\oplus}}{\mu_{5}^{3}}
K_{E}(\mu_{5}R_{\oplus})
\left(1 + \mu_{5}r\right)\left(\frac{e^{-\mu_{5}r}}{r^{2}}\right)\hat{r}
\nonumber \\
&=\frac{f_{5}}{4\pi}\frac{3B_{\oplus}}{\mu_{5}^{3}R_{\oplus}^{3}}
K_{E}(\mu_{5}R_{\oplus})
\left(1 + \mu_{5}r\right)\left(\frac{e^{-\mu_{5}r}}{r^{2}}\right)\hat{r},
\label{E_field}
\\
\vec{\mathscr{B}}_{5}(\vec{r}) &=\frac{f_{5}\rho_{5,\oplus}\omega_{\oplus}}{\mu_{5}^{5}} K_{B}(\mu_{5}R_{\oplus})\left(\frac{e^{-\mu_5r}}{r^{3}}\right)\left[
	2(1 + \mu_{5}r)\cos\theta\,\hat{r}
	+ (1 + \mu_{5}r+ \mu_{5}^{2}r^{2}) \sin\theta\,\hat{\theta}\right]
\nonumber \\
&=\frac{f_{5}}{4\pi}\frac{3B_{\oplus}\omega_{\oplus}}{\mu_{5}^{5}R_{\oplus}^{3}} K_{B}(\mu_{5}R_{\oplus})\left(\frac{e^{-\mu_5r}}{r^{3}}\right)\left[
	2(1 + \mu_{5}r)\cos\theta\,\hat{r}
	+ (1 + \mu_{5}r+ \mu_{5}^{2}r^{2}) \sin\theta\,\hat{\theta}\right].
\label{B_field}
\end{align}
As shown in \ref{Potentials Fields Appendix}, these fields reduce to the analogous electromagnetic results in the massless limit.

With terrestrial experiments in mind, let us now  investigate the fields close to the Earth's surface.  Setting $r = R_{\oplus} + h$, where $h \ll R_{\oplus},$ the fields to first order in $h/R_{\oplus}$ are
\begin{eqnarray}
\vec{\mathscr{E}}_5(r \simeq R_{\oplus}) 
&\simeq& \frac{f_{5}}{4\pi}\,\frac{B_{\oplus}}{R_{\oplus}^{2}} \left[\frac{3}{\mu_{5}^{3}R_{\oplus}^{3}}
K_{E}(\mu_{5}R_{\oplus})e^{-\mu_{5}R_{\oplus}}\right]
\left[1 - \frac{2h}{R_{\oplus}} + \mu R_{\oplus}\left(1 - \frac{h}{R_{\oplus}}\right)\right]
e^{-\mu_{5} h}\,\hat{r},
\nonumber \\
\label{Earth Surface General E5}
\\
\vec{\mathscr{B}}_{5}(r \simeq R_{\oplus}) 
&\simeq& \frac{f_{5}}{4\pi}\,\frac{B_{\oplus}}{R_{\oplus}^{2}}\, \left[\frac{3\omega_{\oplus}R_{\oplus}}{\mu_{5}^{5}R_{\oplus}^{5}} K_{B}(\mu_{5}R_{\oplus})e^{-\mu_{5}R_{\oplus}}\right]e^{-\mu_{5} h}
\nonumber \\
&&\mbox{} \times
\left\{
	2\left[1 - \frac{3h}{R_{\oplus}} + \mu_{5}R_{\oplus}\left(1 - \frac{2h}{R_{\oplus}}\right)\right]\cos\theta\,\hat{r} \right.
\nonumber \\
&& \left.	+ \left[1 - \frac{3h}{R_{\oplus}} + \mu_{5}R_{\oplus}\left(1 - \frac{2h}{R_{\oplus}}\right) + \mu_{5}^{2}R_{\oplus}^{2}\left(1 - \frac{h}{R_{\oplus}}\right)\right] \sin\theta\,\hat{\theta}\right\}.
\label{Earth Surface General B5}
\end{eqnarray}
For a long-range 5th force, $\mu_{5}R_{\oplus} \ll 1$ (i.e., $\lambda_{5} \gg R_{\oplus}$), so the fields close to the Earth's surface are approximately uniform and are given by
\begin{align}
\vec{\mathscr{E}}_5(r = R_{\oplus}) 
 &\simeq \frac{f_{5}}{4\pi}\frac{B_{\oplus}}{R_{\oplus}^{2}}\,\hat{r},
 \label{E_approx}
\\
\vec{\mathscr{B}}_{5}(r = R_{\oplus}) 
&\simeq\frac{f_{5}}{4\pi}\frac{B_{\oplus}\omega_{\oplus}}{5R_{\oplus}}
	\left(2\cos\theta\,\hat{r}
	+  \sin\theta\,\hat{\theta}\right).
\label{B_approx}
\end{align}
For a short-range 5th force, which is of more interest here, $\mu_{5}R_{\oplus} \gg 1$ (i.e., $\lambda_{5} \ll R_{\oplus}$). Then the   fields close to the Earth's surface are given by
\begin{eqnarray}
\vec{\mathscr{E}}_5(r \simeq R_{\oplus}) 
&\simeq& \frac{1}{2}f_{5}\rho_{5,\oplus}\left(1 - \frac{h}{R_{\oplus}}\right)\frac{e^{-\mu_{5}h}}{\mu_{5}}
\,\hat{r},
\label{Earth Surface short E5}
\\
\vec{\mathscr{B}}_{5}(r \simeq R_{\oplus}) 
&\simeq& \frac{1}{2}f_{5}\rho_{5,\oplus}\left(\frac{\omega_{\oplus}R_{\oplus}\sin\theta}{\mu_{5}}\right)\left(1 - \frac{2h}{R_{\oplus}}\right)e^{-\mu_{5}h}\,\hat{\theta}.
\label{Earth Surface short B5}
\end{eqnarray}
To see that these results are what one should expect, we recognize that on the surface of the Earth ($h = 0$) the fields are only due to the baryons within range $\lambda_{5}$ of a point on the surface, $\rho_{5,\oplus}(2\pi\lambda_{5}^{3}/3)$.  Furthermore, the magnetic field will be due to the circular motion of those baryons, which are moving with speed $\omega_{\oplus}R_{\oplus}\sin\theta$.  Therefore,
\begin{eqnarray}
|\vec{\mathscr{E}}_5(r = R_{\oplus})|
&\sim& \frac{f_{5}}{4\pi} \frac{\rho_{5,\oplus}(2\pi\lambda_{5}^{3}/3)}{\lambda^{2}}
\sim f_{5}\rho_{5,\oplus}\lambda_{5} = \frac{f_{5}\rho_{5,\oplus}}{\mu_{5}},
\\
|\vec{\mathscr{B}}_{5}(r = R_{\oplus})|
&\sim & \frac{f_{5}}{4\pi} \frac{\rho_{5,\oplus}(2\pi\lambda_{5}^{3}/3)}{\lambda^{2}}\omega_{\oplus}R_{\oplus}\sin\theta
\sim  f_{5}\rho_{5,\oplus}\lambda_{5}\omega_{\oplus}R_{\oplus}\sin\theta \nonumber \\
&=& f_{5}\rho_{5,\oplus}\left(\frac{\omega_{\oplus}R_{\oplus}\sin\theta}{\mu_{5}}\right).
\end{eqnarray}
Thus, it is more natural to express the fields for a short-ranged 5th force in terms of the baryon density $\rho_{5,\oplus}$ rather than the total number of baryons of the Earth $B_{\oplus}$.

\subsection{5th Force Fields Relative to the Surface of the Rotating Earth}

The fields derived in the previous section would be those observed by an inertial observer hovering above the rotating Earth.  However, terrestrial experiments are performed on the Earth's surface, which is a rotating reference frame.  Following the results of electrodynamics in rotating frames \cite{McDonald,Speake}, if $\vec{\mathscr{E}}_5(\vec{r})$ and $\vec{\mathscr{B}}_5(\vec{r})$ are the fields in an inertial frame, then the corresponding fields in a nonrelativistic frame rotating with angular velocity $\vec{\Omega}$ are
\begin{table}[t]
\tbl{Relations between inertial spherical coordinates and the approximately Cartesian coordinates on the surface of the rotating Earth, appropriate for laboratory experiments. It is assumed that $h = z' \ll R_{\oplus}$. (See also Fig.~\ref{fig:Earth}.) }
{\begin{tabular}{@{}c|l@{}} \toprule
  Inertial Spherical Coordinates & Surface Cartesian Coordinates \\ \hline
       $\hat{r}$ & $\hat{z}'$ (vertically upward) \\
       $\hat{\theta}$ & $\hat{y}'= \hat{\varphi} = -\hat{\theta}$ (north)\\
              $\hat{\phi}$ & $\hat{x}'$ (east) \\
          $r = R_{\oplus} + h$ & $z' = h$ (height above the surface) \\
               $\theta$ & $\varphi = \pi/2 - \theta$  (latitude) \\
                       \botrule
   \end{tabular}     \label{tab:Coordinates}}
\end{table}
%
\label{rotating fields}
\begin{align}
    \vec{\mathscr{E}}_{5}'(\vec{r}) &\simeq \vec{\mathscr{E}}_5(\vec{r}) + (\vec{\Omega} \times \vec{r}) \times  \vec{\mathscr{B}}_{5}(\vec{r}),
\label{rotating E}
    \\
    \vec{\mathscr{B}}_{5}'(\vec{r}) & \simeq \vec{\mathscr{B}}_{5}(\vec{r}).
    \label{rotating B}
\end{align}
%
Eqs.~(\ref{rotating E}) and (\ref{rotating B}) are consistent with the fact that the force on a charged (nonrelativistic) particle is independent of reference frame.  As viewed in the inertial frame, the 5th force on a nucleon with coupling  $f_{5}$ at rest in the rotating frame has contributions from both the hyperelectric and hypermagnetic fields, 
\begin{equation}
    \vec{F}_{5}(\vec{r}) = f_{5}\left[\vec{\mathscr{E}}_5(\vec{r}) + \vec{v}\times \vec{\mathscr{B}}_5(\vec{r})\right],
\end{equation}
where $\vec{v} = \vec{\Omega} \times \vec{r}$.  On the other hand, in the rotating frame, the 5th force on the nucleon must be due entirely to the hyperelectric field since the nucleon is at rest ($\vec{v}\,' = 0$):
\begin{equation}\label{static_force}
    \vec{F}'_{5}(\vec{r}\,') = f_{5}\vec{\mathscr{E}}'_5(\vec{r}\,').
\end{equation}
Since $\vec{F}_{5} = \vec{F}'_{5}$, Eq.~(\ref{rotating E}) follows immediately.

To obtain the hyperelectric and hypermagnetic fields observed on the surface of the Earth, we first note that $\vec{\Omega} = \omega_{\oplus}\hat{z}$. Then, near  the Earth's surface, 
\begin{equation}
    \vec{\Omega}\times \vec{r} \simeq (R_{\oplus}+ h)\omega_{\oplus}\sin\theta\,\hat{\phi},
\end{equation} 
so 
\begin{align}
(\vec{\Omega}\times \vec{r})\times \hat{r} &\simeq (R_{\oplus}+h)\omega_{\oplus}\sin\theta\,\hat{\theta},
\label{r cross product}
\\
(\vec{\Omega}\times \vec{r})\times \hat{\theta} &\simeq -(R_{\oplus} + h)\omega_{\oplus}\sin\theta\,\hat{r}.
\label{theta cross product}
\end{align}
Using these results, we find that the hyperelectric field  observed on the surface of the Earth  has an additional contribution from the inertial hypermagnetic field.  However, for experiments conducted on the Earth's surface, it is convenient to re-express the result in terms of an approximately Cartesian set of coordinates with the axes $\hat{x}'$, $\hat{y}'$, and $\hat{z}'$ as shown in Fig.~\ref{fig:Earth}, and replace the polar angle $\theta$ by the latitude $\varphi$ (Table~\ref{tab:Coordinates}).  Combining Eqs.~(\ref{Earth Surface General E5}), (\ref{Earth Surface General B5}),  (\ref{rotating E}), (\ref{r cross product}), and (\ref{theta cross product}) with these new coordinates, the hyperelectric field observed on the Earth to first order in $z'/R_{\oplus}$ is
\begin{eqnarray}
 \vec{\mathscr{E}}'_5(r \simeq R_{\oplus}) &=&\frac{f_{5}}{4\pi}\,\frac{B_{\oplus}}{R_{\oplus}^{2}} \left[\frac{3}{\mu_{5}^{3}R_{\oplus}^{3}}
K_{E}(\mu_{5}R_{\oplus})e^{-\mu_{5}R_{\oplus}}\right]
\left[1 - \frac{2z'}{R_{\oplus}} + \mu R_{\oplus}\left(1 - \frac{z'}{R_{\oplus}}\right)\right]
e^{-\mu_{5} z'}\,\hat{z}'
\nonumber \\
&& \mbox{}  -\frac{f_{5}}{4\pi}\,\frac{B_{\oplus}}{R_{\oplus}^{2}}\left[\frac{3\omega_{\oplus}^{2}R_{\oplus}^{2}}{\mu_{5}^{5}R_{\oplus}^{5}}K_{B}(\mu_{5}R_{\oplus})e^{-\mu_{5}R_{\oplus}}\right]e^{-\mu_{5} z'}
\nonumber \\
&& \mbox{} \times \left\{
\left[1 - \frac{2z'}{R_{\oplus}} + \mu_{5}R_{\oplus}\left(1 - \frac{z'}{R_{\oplus}}\right)\right]
\sin 2\varphi\,\,\hat{y}'
\right.
\nonumber \\
&& \mbox{} + \left. 
\left[1 - \frac{2z'}{R_{\oplus}} + \mu_{5}R_{\oplus}\left(1 - \frac{z'}{R_{\oplus}}\right) + \mu_{5}^{2}R_{\oplus}^{2}\right]
\cos^{2}\varphi\,\,\hat{z}'
\right\}.
\label{Earth Surface Rotating E5}
\end{eqnarray}
We see that the vertical component of the hyperelectric field is slightly decreased by the effects of the rotating frame.  In addition, there is now a new small horizontal component directed toward the south that does not arise in treatments where the Earth's rotation has not been taken into account.

As shown in Eq.~(\ref{rotating B}), the form of the hypermagnetic field remains unchanged from its inertial value.  Changing to the approximately Cartesian coordinates given in Table~\ref{tab:Coordinates}, the hypermagnetic field to first order in $z'/R_{\oplus}$ is given by
\begin{eqnarray}
\vec{\mathscr{B}}_{5}'(r \simeq R_{\oplus}) 
&\simeq& \frac{f_{5}}{4\pi}\,\frac{B_{\oplus}}{R_{\oplus}^{2}}\, \left[\frac{3\omega_{\oplus}R_{\oplus}}{\mu_{5}^{5}R_{\oplus}^{5}} K_{B}(\mu_{5}R_{\oplus})e^{-\mu_{5}R_{\oplus}}\right]e^{-\mu_{5} z'}
\nonumber \\
&&\mbox{} \times
\left\{
	2\left[1 - \frac{3z'}{R_{\oplus}} + \mu_{5}R_{\oplus}\left(1 - \frac{2z'}{R_{\oplus}}\right)\right]\sin\varphi\,\,\hat{z}' \right.
\nonumber \\
&& \left.	- \left[1 - \frac{3z'}{R_{\oplus}} + \mu_{5}R_{\oplus}\left(1 - \frac{2z'}{R_{\oplus}}\right) + \mu_{5}^{2}R_{\oplus}^{2}\left(1 - \frac{z'}{R_{\oplus}}\right)\right] \cos\varphi\,\,\hat{y}'\right\}. \nonumber \\
\label{Earth Surface Rotating B5}
\end{eqnarray}
The magnitude of the vertical component of the hypermagnetic field is maximized at the poles ($\varphi = \pm\pi/2$), while the horizontal component is maximized at the equator ($\varphi = 0$).

We can now take another look at limiting cases of long- and short-ranged 5th forces, this time as viewed relative to the laboratory fixed to the rotating Earth, instead of being relative to an inertial observer as in Eqs.~(\ref{E_approx})--(\ref{Earth Surface short B5}).  When $\mu_{5}R_{\oplus} \ll 1$ (i.e., $\lambda_{5} \gg R_{\oplus}$), we then have near the Earth's surface,
\begin{align}
\vec{\mathscr{E}}_{5}'(r \simeq R_{\oplus}) 
 &\simeq \frac{f_{5}}{4\pi}\frac{B_{\oplus}}{R_{\oplus}^{2}}\left(1 - \frac{2z'}{R_{\oplus}}\right)
 \left[\left(1 - \frac{\omega_{\oplus}^{2}R_{\oplus}^{2}}{5}\cos^{2}\varphi\right)\hat{z}'
 - \frac{\omega_{\oplus}^{2}R_{\oplus}^{2}}{5}\sin 2\varphi\,\,\hat{y}' \right],
\label{Rotating long range E5}
\\
\vec{\mathscr{B}}_{5}'(r \simeq R_{\oplus}) 
&\simeq\frac{f_{5}}{4\pi}\frac{B_{\oplus}\omega_{\oplus}}{5R_{\oplus}}\left(1 - \frac{3z'}{R_{\oplus}}\right)
	\left(2\sin\varphi\,\hat{z}'
	-  \cos\varphi\,\hat{y}'\right),
\label{Rotating long range B5}
\end{align}
while for the short-ranged 5th force,  $\mu_{5}R_{\oplus} \gg 1$ (i.e., $\lambda_{5} \ll R_{\oplus}$), we have
\begin{eqnarray}
\vec{\mathscr{E}}_5(r \simeq R_{\oplus}) 
&\simeq& \frac{1}{2}f_{5}\rho_{5,\oplus}\frac{e^{-\mu_{5}z'}}{\mu_{5}}
\left[\left(1 - \omega_{\oplus}^{2}R_{\oplus}^{2}\cos^{2}\varphi\right)\hat{z}'
- \left(\frac{\omega_{\oplus}^{2}R_{\oplus}^{2}\sin 2\varphi}{\mu_{5}R_{\oplus}}\right)\hat{y}'\right],
\label{Rotating short range E5}
\\
\vec{\mathscr{B}}_{5}'(r \simeq R_{\oplus}) 
&\simeq& -\frac{1}{2}f_{5}\rho_{5,\oplus}\left(\frac{\omega_{\oplus}R_{\oplus}\cos\varphi}{\mu_{5}}\right)e^{-\mu_{5}z'}\,\hat{y}'.
\label{Rotating short range B5}
\end{eqnarray}

 \subsection{Units}
 
Up to this point we have used natural units where $\hbar = c = 1$ for convenience in the theoretical analysis. Just as there are a number of different conventions for units and dimensions in classical electrodynamics (e.g., SI, Gaussian, and Heaviside-Lorentz), there is no unique choice for a classical vector field 5th force.  Since most experimental measurements are now reported in SI units, it would be most convenient to insert factors of $\hbar$ and $c$ so that resulting quantities have simple SI units.  Until now, our treatment has generally followed the conventions of naturalized electromagnetic Heaviside-Lorentz units, which are most commonly used in fundamental physics.  In electromagnetism, the electromagnetic quantities in this convention do not have named units, which makes evaluating and reporting experimental quantities awkward.  The same issues apply to the 5th force here.  

In order for the 5th force quantities to correspond as closely as possible to those of Heaviside-Lorentz electromagnetism, we will insert the factors of $\hbar$ and $c$ in the field equations of a vector field 5th force using the replacements
\begin{subequations}
\begin{align}
    f_{5} &\rightarrow \sqrt{\hbar c}\,f_{5}, \\
    \mu_{5} &\rightarrow \frac{\mu_{5}c}{\hbar},\\
     J^{\mu}_{5} = (\rho_{5},\vec{J}_{5}) &\rightarrow J^{\mu}_{5} = (c\rho_{5},\vec{J}_{5}).
\end{align}
\end{subequations}
In addition, all velocities are written in terms of $c$: $\vec{v} \rightarrow \vec{v}/c.$    Using these results, the force on a nucleon moving with velocity $\vec{v}$ in the 5th force fields $\vec{\mathscr{E}}_{5}$ and $\vec{\mathscr{B}}_{5}$ is given by
\begin{equation}
    \vec{F} = \sqrt{\hbar c}\,f_{5}\left(\vec{\mathscr{E}}_{5} + \frac{\vec{v}}{c} \times \vec{\mathscr{B}}_{5}\right).
\end{equation}
The factors of $\hbar$ and $c$ can be hidden in all expressions if, as in Heaviside-Lorentz electromagnetism, we  replace the dimensionless nucleon coupling constant $f_{5}$ with the dimensionful charge
\begin{equation}
    \tilde{f}_{5}\equiv \sqrt{\hbar c}\,f_{5},
    \end{equation}
which has dimensions $(\mbox{energy}\cdot\mbox{length})^{1/2}$.  The hyperelectric and hypermagnetic fields then have dimensions $\mbox{force}/\mbox{charge} =(\mbox{energy}/\mbox{volume})^{1/2}$. Therefore, in SI units, the charge would have units ${\rm (J\cdot m)^{1/2} = (N\cdot m^{2})^{1/2}}$, and the fields would have units ${\rm (J/m^{3})^{1/2} = (N/m^{2})^{1/2}}$.

\section{Terrestrial 5th Force Experiments}

\subsection{Formalism for 5th Force Experiments}

As discussed in the Introduction, a large and continuing experimental effort has set stringent limits on new Yukawa and inverse-square-law forces over a wide range of distances and couplings.  This includes a 5th force coupling to baryon number.  Many of these experiments use the Earth as a source of the 5th force, but they are only searching for the simplest radial component of the hyperelectric field.  As we have seen, the 4-vector nature of the 5th force gives rise to new contributions to the radial and tangential components for experiments conducted on the surface arising from the Earth's rotation.  In this section, we will briefly review some of the 5th force experiments in regards to their sensitivity to these new contributions.  In particular, this could be important for a scalar-vector 5th force as described in \ref{Scalar-Vector Appendix}.  

Since the 5th force phenomenology applied to terrestrial experiments usually parameterizes the interaction relative to gravity, we will begin by reformulating the formulas for the total force acting on a nucleon in terms which will facilitate comparison with existing limits on the 5th force from these experiments.  Consider the total force $ \vec{F}_{i}$ (gravitational and 5th force) acting on a small test body of mass $m_{i}$ and baryon number $B_{i}$ moving with velocity $\vec{v}_{i}\,'$ relative to the Earth's surface which is given by
\begin{equation}
    \vec{F}_{i} = -m_{i}g\,\hat{z}' + \sqrt{\hbar c}f_{5}B_{i}\left(\vec{\mathscr{E}}_{5}' + \frac{\vec{v}_{i}\,'}{c} \times \vec{\mathscr{B}}_{5}'\right).
    \label{F total 1}
\end{equation}
Here $g = |\vec{g}|$ is the magnitude of the local gravitational field $\vec{g}$, which is given by
\begin{equation}
    \vec{g} = -\frac{GM_{\oplus}}{R_{\oplus}^{2}}\hat{z}',
\end{equation}
and $\vec{\mathscr{E}}_{5}'$ and $\vec{\mathscr{B}}_{5}'$ are the local hyperelectric and hypermagnetic fields arising from the Earth, as presented in Eqs.~(\ref{Earth Surface Rotating E5}) and (\ref{Earth Surface Rotating B5}), respectively.  If we now express the masses in terms of the mass of hydrogen $m_{\rm H}$, Eq.~(\ref{F total 1}) can be recast into the form
\begin{equation}
    \vec{F}_{i} = -m_{i}g\left[\hat{z}' - \frac{\sqrt{\hbar c}f_{5}}{Gm_{\rm H}^{2}}\frac{R_{\oplus}^{2}}{\mu_{\oplus}}\left(\frac{B_{i}}{\mu_{i}}\right)\left(\vec{\mathscr{E}}_{5}' + \frac{\vec{v}_{i}\,'}{c} \times \vec{\mathscr{B}}_{5}'\right)\right],
    \label{F total 2}
\end{equation}
where $\mu_{i} = m_{i}/m_{\rm H}$ and $\mu_{\oplus} = m_{\oplus}/m_{\rm H}$.  For the 5th force contribution, let us now factor out the hyperelectric field strength of the non-rotating Earth for the massless boson,
\begin{equation}
    \mathscr{E}_{5,\mu_{5} = 0} = \frac{\sqrt{\hbar c}f_{5}}{4\pi}\frac{B_{\oplus}}{R_{\oplus}^{2}}, 
\end{equation}
which gives
\begin{equation}
    \vec{F}_{i} = -m_{i}g\left[\hat{z}' - \frac{\hbar c f_{5}^{2}}{4\pi Gm_{\rm H}^{2}}\left(\frac{B_{i}}{\mu_{i}}\right)\left(\frac{B_{\oplus}}{\mu_{\oplus}}\right)\left(\frac{\vec{\mathscr{E}}_{5}'}{\mathscr{E}_{5,\mu_{5} = 0}} + \frac{\vec{v}_{i}\,'}{c} \times \frac{\vec{\mathscr{B}}_{5}'}{\mathscr{E}_{5,\mu_{5} = 0}}\right)\right].
    \label{F total 3}
\end{equation}
Eq.~(\ref{F total 3}) can be rewritten in terms of the dimensionless parameters $\alpha_{i,\oplus}$ and $\xi_{5}$,
\begin{equation}
    \vec{F}_{i} = -m_{i}g\left[\hat{z}' - \alpha_{i,\oplus}\left(\frac{\vec{\mathscr{E}}_{5}'}{\mathscr{E}_{5,\mu_{5} = 0}} + \frac{\vec{v}_{i}\,'}{c} \times \frac{\vec{\mathscr{B}}_{5}'}{\mathscr{E}_{5,\mu_{5} = 0}}\right)\right],
    \label{F total 4}
\end{equation}
where 
\begin{equation}
    \alpha_{i,\oplus} \equiv \xi_{5}\left(\frac{B_{i}}{\mu_{i}}\right)\left(\frac{B_{\oplus}}{\mu_{\oplus}}\right),
\end{equation}
is approximately the strength of the 5th force relative to gravity, 
\begin{equation}
    \xi_{5} \equiv \frac{f_{5}^{2}}{4\pi}\left(\frac{m_{P}}{m_{\rm H}}\right)^{2},
\end{equation}
and $m_{P} = (\hbar c/G)^{1/2}$ is the Planck mass.

Most laboratory experiments are interested in constraining new forces with ranges $\lambda_{5} \ll R_{\oplus}$.  Using Eqs.~(\ref{Rotating short range E5}) and (\ref{Rotating short range B5}), the 5th force fields in this limit are given by
\begin{align}
    \frac{\vec{\mathscr{E}}_{5}'}{\mathscr{E}_{5,\mu_{5} = 0}} &\simeq
    \frac{3}{2}\frac{e^{-\mu_{5}z'}}{\mu_{5}R_{\oplus}}
\left[\left(1 - \frac{\omega_{\oplus}^{2}R_{\oplus}^{2}}{c^{2}}\cos^{2}\varphi\right)\hat{z}'
- \left(\frac{\omega_{\oplus}^{2}R_{\oplus}^{2}}{c^{2}}\right)\left(\frac{\sin 2\varphi}{\mu_{5}R_{\oplus}}\right) \hat{y}'\right],
\label{short-range E' ratio}
\\
\frac{\vec{\mathscr{B}}_{5}'}{\mathscr{E}_{5,\mu_{5} = 0}} &\simeq
-\frac{3}{2}\frac{e^{-\mu_{5}z'}}{\mu_{5}R_{\oplus}}\left(\frac{\omega_{\oplus}R_{\oplus}}{c}\right)\cos\varphi\,\hat{y}'.
\label{short-range B' ratio}
\end{align}
Substituting Eqs.~(\ref{short-range E' ratio}) and (\ref{short-range B' ratio}) back into Eq.~(\ref{F total 4}) then gives,   for a short-ranged 5th force,
\begin{eqnarray}
    \vec{F}_{i} &\simeq & -m_{i}g\left\{\hat{z}' - \alpha_{i,\oplus}
    \left(\frac{3}{2}\frac{e^{-\mu_{5}z'}}{\mu_{5}R_{\oplus}}\right)
    \left[\left(1 - \frac{\omega_{\oplus}^{2}R_{\oplus}^{2}}{c^{2}}\cos^{2}\varphi\right)\hat{z}'
- \left(\frac{\omega_{\oplus}^{2}R_{\oplus}^{2}}{c^{2}}\right)\left(\frac{\sin 2\varphi}{\mu_{5}R_{\oplus}}\right) \hat{y}'\right.\right.
\nonumber \\
&& \left.\left.\mbox{} -\left(\frac{\omega_{\oplus}R_{\oplus}}{c}\cos\varphi\right) \left(\frac{\vec{v}_{i}\,'}{c} \times \hat{y}'\right)\right]\right\}.
    \label{short-range F total}
\end{eqnarray}
For convenience, the numerical values of terrestrial parameters for evaluating experimental results are given in Table~\ref{tab:terrestrial parameters}.  It is important to note, however, that for short-ranged 5th forces, the local matter distribution will provide a significant contribution to the force on a test body beyond the simple uniform spherical Earth model discussed here.  These additional contributions are beyond the scope of the present paper.

\begin{table}[t]
\tbl{Numerical Values of Terrestrial Parameters}
{\begin{tabular}{@{}l|l@{}} \toprule
  Parameter & Value\\ \hline
         Mass  & $M_{\oplus} = 5.97 \times 10^{24}$~kg \\
         Radius & $R_{\oplus} = 6.38 \times 10^{6}$~m \\ 
         Baryon Number & $ B_{\oplus} =  3.57 \times 10^{51}$ \\
         Baryon Density & $\rho_{5,\oplus} = 3.28 \times 10^{30}$~m$^{-3}$ \\
         Baryon-to-Mass Ratio & $B_{\oplus}/\mu_{\oplus} \simeq 1$ \\
          Equatorial Speed & $\displaystyle \frac{\omega_{\oplus}R_{\oplus}}{c} = 1.55 \times 10^{-6}$   \\                    \botrule
   \end{tabular}     \label{tab:terrestrial parameters}}
\end{table}

All previous 5th force experiments using the Earth as a source have focused entirely on searching for a static hyperelectric 5th force, which assumes $\omega_{\oplus} =0$ \cite{Fischbach_book}:
\begin{equation}
    \vec{F}_{i,\omega_{\oplus} = 0} \simeq  -m_{i}g\left[1 - \alpha_{i,\oplus}
    \left(\frac{3}{2}\frac{e^{-\mu_{5}z'}}{\mu_{5}R_{\oplus}}\right)
   \right]\hat{z}'.
    \label{short-range hyperelectric total}
\end{equation}
This is expected since  $\omega_{\oplus}R_{\oplus}/c \sim 10^{-6}$, so the hypermagnetic force is suppressed relative to  the hyperelectric force.  However, the stringent constraints on $\alpha_{i,\oplus}$ obtained using this assumption would not apply to the scalar-vector 5th force scenario described in \ref{Scalar-Vector Appendix}.  In this case, the $\omega_{\oplus}$-independent vector 5th force would be cancelled by the scalar 5th force, leaving the purely hypermagnetic 5th force.  For the short-ranged scalar-vector 5th force, the total 5th force acting on a test mass is then due only to the hypermagnetic 5th force,  
\begin{eqnarray}
    \vec{F}_{i}^{\rm mag} &\simeq & - \alpha_{i,\oplus}^{\rm SV}m_{i}g \left(\frac{3}{2}\frac{e^{-\mu_{5}z'}}{\mu_{5}R_{\oplus}}\right)
    \left[\left(\frac{\omega_{\oplus}^{2}R_{\oplus}^{2}}{c^{2}}\cos^{2}\varphi\right)\hat{z}'
+ \left(\frac{\omega_{\oplus}^{2}R_{\oplus}^{2}}{c^{2}}\right)\left(\frac{\sin 2\varphi}{\mu_{5}R_{\oplus}}\right) \hat{y}'\right.
\nonumber \\
&& \left.\mbox{} +\left(\frac{\omega_{\oplus}R_{\oplus}}{c}\cos\varphi\right) \left(\frac{\vec{v}_{i}\,'}{c} \times \hat{y}'\right)\right].
    \label{short-range SV}
\end{eqnarray}

We see from Eq.~(\ref{short-range SV}) that the magnetic 5th force has three distinct contributions within the square brackets characterized by different directions and latitude dependence. 
This contrasts with the static gravitational force and the 5th force with   $\omega_{\oplus} =0$ given by Eq.~(\ref{short-range hyperelectric total}), which are purely vertical locally in this simple model, and independent of location.
The first term in square brackets of Eq.~(\ref{short-range SV}),
\begin{equation}
     \vec{F}_{i,z}^{\rm mag} \simeq  - \alpha_{i,\oplus}^{\rm SV}m_{i}g \left(\frac{3}{2}\frac{e^{-\mu_{5}z'}}{\mu_{5}R_{\oplus}}\right)
    \left(\frac{\omega_{\oplus}^{2}R_{\oplus}^{2}}{c^{2}}\cos^{2}\varphi\right)\hat{z}',
    \label{F mag z}
\end{equation}
is purely vertical, directed downward parallel to the local gravitational field.  The latitude dependence, given by $\cos^{2}\varphi$, is maximized at the equator and vanishes at the poles.  The second term in square brackets of Eq.~(\ref{short-range SV}),
\begin{equation}
     \vec{F}_{i,y}^{\rm mag} \simeq  - \alpha_{i,\oplus}^{\rm SV}m_{i}g \left(\frac{3}{2}\frac{e^{-\mu_{5}z'}}{\mu_{5}R_{\oplus}}\right)
    \left(\frac{\omega_{\oplus}^{2}R_{\oplus}^{2}}{c^{2}}\right)\left(\frac{\sin 2\varphi}{\mu_{5}R_{\oplus}}\right) \hat{y}',
    \label{F mag y}
\end{equation}
is directed horizontally locally toward the equator since $\sin 2\varphi < 0$ in the southern hemisphere, and is maximized at $\pm 45^{\circ}$ latitude.  The remaining term in the square brackets of Eq.~(\ref{short-range SV}) represents the velocity-dependent magnetic 5th force,
\begin{equation}
     \vec{F}_{i,v}^{\rm mag} \simeq  - \alpha_{i,\oplus}^{\rm SV}m_{i}g \left(\frac{3}{2}\frac{e^{-\mu_{5}z'}}{\mu_{5}R_{\oplus}}\right)
    \left(\frac{\omega_{\oplus}R_{\oplus}}{c}\cos\varphi\right) \left(\frac{\vec{v}_{i}\,'}{c} \times \hat{y}'\right).
    \label{F mag v}
\end{equation}
$\vec{F}_{i,v}^{\rm mag}$  has a complicated directional dependence depending on the velocity of the test mass, but it has its maximum magnitude on the equator and vanishes at the poles. 

Finally, it is useful to compare the 5th force magnetic force components given by Eqs.~(\ref{F mag z})--(\ref{F mag v}) with the inertial forces experienced by a test body on the surface of the Earth, which have similar form.  The centrifugal and Coriolis forces on the test mass $m_{i}$ are
\begin{equation}
    \vec{F}_{i}^{\rm cent}= m\omega_{\oplus}^{2}R_{\oplus}\left(\cos^{2}\varphi\,\hat{z}' - \frac{1}{2}\sin 2\varphi\,\hat{y}'\right),
\end{equation}
and
\begin{equation}
    \vec{F}_{i}^{\rm Cor}= 2m\omega_{\oplus}\left[\sin\varphi (\vec{v}\,' \times \hat{z}')
    + \cos\varphi(\vec{v}\,'\times\hat{y}')\right],
\end{equation}
respectively.

In the next subsections, we briefly discuss the phenomenological consequences of the magnetic 5th force for terrestrial 5th force experiments.

\subsection{Composition-Independent Experiments}

It is customary to classify 5th force experiments based on whether or not they depend on the compositions of the test masses used \cite{Fischbach_book}.  For composition-independent experiments, it is most common to write the total interacting potential between two particles with masses $m_{1}$ amd $m_{2}$ separated by a distance $r$ as
\begin{equation}
    V(r) = -\frac{Gm_{1}m_{2}}{r}\left(1 + \alpha_{5} e^{-r/\lambda_{5}}\right),
    \label{V composition-independent}
\end{equation}
where $\alpha_{5}$ is a universal constant measuring the strength of the 5th force relative to gravity, and $\lambda_{5}$ is the range.  For a short-ranged 5th force using the Earth as a source, Eq.~(\ref{V composition-independent}) corresponds to Eq.~(\ref{short-range hyperelectric total}) where $\alpha_{5} = \alpha_{i,\oplus}.$  The hypermagnetic forces given by Eqs.~(\ref{F mag z})--(\ref{F mag v}) have similar $z'$-dependence, but have location and directional differences.

Composition-independent experiments  constrain $\alpha_{5}$ as a function of $\lambda_{5}$ by searching for violations of the inverse-square-law (ISL) of Newtonian gravity \cite{Fischbach_book,Adelberger ISL,Newman ISL}.  These experiments are most sensitive when the separation of the test bodies $r \sim \lambda_{5}$.  Terrestrial experiments searching for violations of the ISL over laboratory scales naturally require the use of small source masses \cite{Moody,Hoskins}.  Larger scale ISL terrestrial experiments are inevitably geophysical in nature, utilizing mine shafts \cite{Holding}, boreholes in ice sheets \cite{Zumberge}, lakes \cite{Muller 1989,Muller 1990,Baldi,Cornaz,Hubler}, and towers \cite{Thomas 1989,Liu,Romaides}.  These experiments require careful surveys of gravitational fields  and/or accurate modeling of subsurface density \cite{Bartlett,Thomas 1990}.  For example, for a tower experiment, one uses gravimeters to measure the gravitational field as a function of height $z'$.  Then the difference between the measured value $g_{\rm measured}(z')$, and the modeled value $g_{\rm model}(z')$ assuming Newtonian gravity is correct, is given by \cite{Romaides}
\begin{equation}
    \Delta g(z) = g_{\rm measured}(z') - g_{\rm model}(z') = 2\pi\rho_{m}G\alpha_{5}\lambda_{5}\left(e^{-z'/\lambda_{5}} -1 \right),
\end{equation}
where $\rho_{m}$ is the average mass density of the soil surrounding the tower.  Presently, the most stringent limits on $\alpha_{5}$ in the $\sim 10$--$1000$~m range, $\alpha_{5}\lesssim 10^{-3}$, come from geophysical experiments that were conducted over twenty years ago \cite{Murata}.

Most of the geophysical 5th force ISL experiments are searching for non-Newtonian gravity effects in the vertical direction.  As seen in Eq.~(\ref{F mag z}), the magnetic 5th force in the vertical direction is suppressed relative to electric 5th force by the rotational factor $\omega_{\oplus}^{2}R_{\oplus}^{2}/c^{2}$.  However, in the scalar-vector 5th force model discussed in \ref{Scalar-Vector Appendix}, the magnetic 5th force contribution becomes dominant.  Since the $z'$-dependence of the vertical component of the magnetic 5th force is the same as for the electric 5th force, one can estimate the limits on $\alpha^{\rm SV}_{i,\oplus}$ when $\lambda_{5} \sim 10$--$1000$~m  from the existing geophysical limits on the composition-independent $\alpha_{5}$.  This gives
\begin{equation}
    \alpha^{\rm SV}_{i,\oplus} \lesssim \alpha_{5} \left(\frac{\omega_{\oplus}R_{\oplus}}{c}\right)^{-2}
    \sim 10^{-3}\left(10^{12}\right) \sim 10^{9}.
\end{equation}

\subsection{Composition-Dependent Experiments}

The second class of 5th force experiments are experiments which depend on the composition of the test bodies and source masses \cite{Fischbach_book}.  Unlike gravity, which couples universally to matter, the other Standard Model interactions and most new forces beyond the Standard Model couple to matter and fields with varying strengths. For example, the 5th force model considered in this paper couples universally to nucleons, but does not couple to leptons and the fundamental gauge bosons.  A signature of  a new long-range composition-dependent force would be a violation of the weak equivalence principle (WEP), so a considerable experimental effort dating back to Newton has searched for violations of the WEP \cite{Will,Tino}.   In this section, we will briefly examine how the effects of a magnetic 5th force might appear in  terrestrial composition-dependent force experiments that use the Earth as the source mass.   Hypermagnetic effects arising in composition-dependent experiments, which use laboratory source masses or local geophysical features such as hills or mountains, are beyond the scope of the present paper.

    \subsubsection{Free-fall Experiments}
    
    The conceptually simplest tests of the WEP compare the free-fall accelerations of two different test bodies falling toward the Earth \cite{Sontag}. If $\vec{a}_{i}$ is the acceleration of the $i$th test body, then the differential acceleration arising from a short-range 5th force given by Eq.~(\ref{short-range F total}) is,
\begin{eqnarray}
    \Delta\vec{\kappa}_{ij} & \equiv & \frac{\vec{a}_{i}' - \vec{a}_{j}'}{g}
    \nonumber \\ 
    &\simeq & 
    \xi_{5}\Delta\left(\frac{B}{\mu}\right)_{ij}
    \left(\frac{3}{2}\frac{\lambda_{5}e^{-z'/\lambda_{5}}}{R_{\oplus}}\right)
    \left[\left(1 - \frac{\omega_{\oplus}^{2}R_{\oplus}^{2}}{c^{2}}\cos^{2}\varphi\right)\hat{z}'
- \left(\frac{\omega_{\oplus}^{2}R_{\oplus}^{2}}{c^{2}}\right)\left(\frac{\lambda_{5}\sin 2\varphi}{R_{\oplus}}\right) \hat{y}'\right].
\nonumber \\
    \label{short-range Delta kappa total}
\end{eqnarray}
  Here we assumed $B_{\oplus}/\mu_{\oplus} \simeq 1$, $v_{i}' \ll \omega_{\oplus}R_{\oplus} \simeq 465$~m/s, and used
  \begin{equation}
    \Delta\left(\frac{B}{\mu}\right)_{ij} \equiv \left(\frac{B_{i}}{\mu_{i}}\right)  - \left(\frac{B_{j}}{\mu_{j}}\right).
  \end{equation}
We see that the magnetic 5th force  slightly reduces the vertical component of the differential acceleration relative to the electric 5th force, while introducing a new horizontal component.  On the other hand, for the scalar-vector 5th force model of \ref{Scalar-Vector Appendix}, the electric 5th force is suppressed relative to the magnetic contribution.  Then the vertical component of the differential acceleration is downward, rather than upward, as if it was attracted to the Earth instead of repelled.  These effects would also depend on latitude, which is not the case for the electric 5th force.  If the horizontal component is significant, it would, in effect, redefine the direction of vertical for the falling body that is composition-dependent, unlike the analogous effect of the centrifugal force which is composition-independent.
    
 By design, free-fall tests of the WEP are most sensitive to the vertical component of a putative 5th force.  The most sensitive experiment using freely falling macroscopic test bodies was done by Kuroda and Mio \cite{Kuroda 1989,Kuroda 1990} who found $\Delta\kappa_{ij}\lesssim 1.0 \times 10^{-10}.$  (See also Ref.~\citen{Sontag} for more recent developments involving drop tower experiments.) Carusotto \textit{et al.} \cite{Carusotto 1992,Carusotto 1993,Carusotto 1996} achieved comparable sensitivity by measuring the rotation of a freely falling disc made of two halves of different composition.  Alternatively, extraordinary progress has been made in atom-interferometric tests of the WEP using freely falling atoms \cite{Tino Atoms}. Most recently, a group at Stanford has found $\Delta\kappa_{ij} \lesssim 10^{-12}$ when comparing the accelerations of $^{85}$Rb and $^{87}$Rb \cite{Asenbaum}.  Constraints on a magnetic 5th force could be obtained by calculating the phase shifts induced on the atoms.

    \subsubsection{Torsion Balance Experiments}
    
Some of the most stringent limits on new forces at laboratory scales come from torsion balance experiments testing the WEP \cite{Adelberger Torsion,Wagner}. This effort began with a series of experiments by  E\"{o}tv\"{o}s, Pek\'{a}r, and Fekete (EPF) \cite{EPF} which achieved the limit $\Delta\kappa_{ij}\lesssim 10^{-9}$ on the differential acceleration of various test bodies towards the Earth.  Later torsion balance experiments  by Roll et al. \cite{Roll}, and Braginskii and Panov \cite{Braginskii}, achieved  $\Delta\kappa_{ij}\lesssim 10^{-11}$ and  $\Delta\kappa_{ij}\lesssim 10^{-12}$, respectively.  However, these experiments studied accelerations toward the Sun, so the results from these experiments cannot set significant limits on a 5th force with range $\lambda_{5} \ll R_{\oplus}.$  Presently, the most stringent test of the WEP from a torsion balance experiment using the Earth as the source is by the E\"{o}t-Wash group, which obtained $\Delta\kappa_{ij}\lesssim 10^{-13}$ \cite{Wagner}.

By design, torsion balances are most sensitive to horizontal forces, which are defined to be those which act perpendicular to the torsion fiber that suspends the test masses.  Even on a perfectly spherical, but rotating Earth, a torsion balance is sensitive to tests of the WEP and a radial hyperelectric 5th force  since the centrifugal force in the rotating frame changes the suspension angle so that it is no longer radial.  From Eq.~(\ref{short-range Delta kappa total}), we see that  $\Delta\kappa_{ij}$ arising from a magnetic 5th force will have a more complicated directional dependence than from the electric 5th force, which has been the expected force in previous experiments. This is especially true for the scalar-vector 5th force model discussed in \ref{Scalar-Vector Appendix}.  Finally, as discussed in Refs.~\citen{Mueterthies} and \citen{Fischbach Moriond}, torsion balances with different configurations of their test masses do not have equal sensitivity to all types of forces.

    \subsubsection{Floating Ball Experiments}
    
We conclude this section on composition-dependent experiments with floating ball experiments \cite{Fischbach_book,Thieberger,Bizzeti 1989,Bizzeti 1990}.  The principle underlying this class of experiments is that when a finite-range composition-dependent force is present, it is nearly impossible for a test body floating in a fluid of different composition from the body to be in equilibrium.  That is, the total buoyant force on the body is not anti-parallel to the body force on the test body.  Because the test body is suspended in a fluid, it is especially sensitive to horizontal forces, as well as vertical forces.  Therefore, such experiments are not only sensitive to short-ranged 5th forces from local vertical geophysical features like cliffs, but also to the horizontal component of a magnetic 5th force of the Earth.  

As noted above, the only published experiment to date which has presented evidence for a 5th force is the floating ball experiment of Thieberger \cite{Thieberger}. In this experiment a copper sphere floating in water at a site near the Hudson river was shown to drift in the direction of the river.   However, another floating ball experiment using a different setup and procedure by Bizzeti  \textit{et al.} found no such effect \cite{Bizzeti 1989,Bizzeti 1990}.  If the systematic issues affecting floating ball experiments \cite{Keyser,Thieberger 1989} can be overcome, it might be useful to repeat improved versions of them because of their sensitivity to forces from arbitrary directions.

\section{Magnetic 5th Force Experiments}

The  5th force experiments discussed in the previous section are designed to detect the effects of a hyperelectric field. As discussed in \ref{Scalar-Vector Appendix}, one can even envision situations where the hyperelectric field is suppressed relative to the hypermagnetic field. Since the vector field model of the 5th force parallels  closely  ordinary electromagnetism, the direct effects of the hypermagnetic  field should appear in experiments analogous to those which are sensitive to the ordinary magnetic field. However, there is one crucial difference: most magnetic experiments involve electrons. In the 5th force model discussed here, leptons do not interact with the hypermagnetic field, hence we will only consider magnetic experiments involving nucleons.    In this section, we will briefly examine two types of experiments which could be sensitive to the magnetic 5th force:  proton cyclotron frequency experiments and  magnetic moment experiments.

       \subsection{Proton Cyclotron Frequency Measurements}
       
   In recent years, a number of groups have been using precision measurements of the cyclotron frequency of charged particles in Penning traps to investigate fundamental physics \cite{Blaum 2010,Dilling,Blaum 2021}.  In these experiments, a strong static magnetic field $\vec{B}$ and a weaker quadrupolar electric field confine the charged particle such that it moves in a plane perpendicular to the magnetic field.  If a 5th force  hypermagnetic field $\vec{\mathscr{B}}_5'$ is also present, the cyclotron frequency $\nu_{c}$ of a trapped proton will be given by
   \begin{equation}
       \nu_{c} = \frac{1}{2\pi} \left(\frac{e|\vec{B}| + f_{5}|\vec{\mathscr{B}}_{5,\perp}'|}{m_{p}}\right),
   \end{equation}
    where $\vec{\mathscr{B}}_{5,\perp}'$ is the  hypermagnetic field perpendicular to the plane of the proton's motion and $m_{p}$ is the mass of the proton. 
    
    Recently, the Baryon Antibaryon Symmetry Experiment (BASE) collaboration used cyclotron frequency measurements to show that the antiproton charge-to-mass ratio was the same as the proton's ratio to 16 parts in $10^{12}$ \cite{Borchert}.  Unfortunately, the extraordinary precision of this result cannot used to constrain a vector 5th force that respects charge-parity-time (CPT) invariance, as in our model.  In this case, both the electric  charge $e$ and the 5th force coupling $f_{5}$ of the antiproton  have opposite signs relative to the proton. Therefore, the ratio of the magnitudes of the antiproton-to-proton cyclotron frequencies, which was measured in the BASE experiment, is unity, irrespective of the magnitude of the 5th force field.  This would not be true for a 5th force which violates CPT.
    
   A change in the magnitude of the cyclotron frequency due to a magnetic 5th force would arise if the direction of the  applied static magnetic field were reversed, $\vec{B} \rightarrow -\vec{B}$.  In this situation, the external hypermagnetic field would not change, but the  sign of the magnetic contribution would change:
   \begin{equation}
       \nu_{c}^{+} = \frac{1}{2\pi} \left(\frac{e|\vec{B}| + f_{5}|\vec{\mathscr{B}}_{5,\perp}'|}{m_{p}}\right)
       \xrightarrow{\vec{B}\rightarrow -\vec{B}}
       \nu_{c}^{-} = \frac{1}{2\pi} \left(\frac{-e|\vec{B}| + f_{5}|\vec{\mathscr{B}}_{5,\perp}'|}{m_{p}}\right).
   \end{equation}
    The resulting change in the magnitude of the cyclotron frequency then depends on $f_{5}$:
    \begin{equation}
        \frac{\nu_{c}^{+}-|\nu_{c}^{-}|}{\nu_{c}^{+}} \simeq \frac{2f_{5}}{e} \left(\frac{|\mathscr{B}_{5,\perp}'|}{|\vec{B}|}\right).
    \end{equation}

    \subsection{Nuclear Spin-Precession Experiments}
  
  Experiments searching for spin-dependent new forces using nuclei should also be sensitive to a magnetic 5th force.  As an example, consider the nuclear spin-precession experiment by Venema \textit{et al.} \cite{Venema}. It was designed to search for a new monopole-dipole potential between two particles \#1 and \#2 of the form   $V_{12}\propto \vec{S}_{1}\cdot\hat{r}_{12}$, where $\vec{S}_{1}$ is the spin of particle \#1 and $\hat{r}_{12}$ is the separation unit vector \cite{Moody Wilcek,Fadeev}.   This was done by measuring the ratio of the nuclear spin-precession frequencies of $^{199}$Hg and $^{201}$Hg for two different orientations of an applied magnetic field $\vec{B}$ relative to the Earth's gravitational field.   They assumed that the source of the scalar interaction was the Earth, while the  Hg nuclei coupling was spin-dependent.  The interaction Hamiltonian for this setup is
  \begin{equation}
      H_{\rm int} = -g_{I}\mu_{N}\vec{I}\cdot\vec{B} + {\cal A}\epsilon\vec{I}\cdot \hat{r}/|\vec{I}|,
      \label{monopole-dipole}
  \end{equation}
  where $g_{I}$ is the nuclear $g$-factor, $\mu_{N}= e\hbar/2m_{p}$ is the nuclear magneton, and $\vec{I}$ is the angular momentum of the nucleus.  Here ${\cal A}$ characterizes the strength of the new interaction, $\hat{r}$ points vertically downward toward the Earth's surface, and $\epsilon \sim 1$ is a model-dependent parameter.  The effect of the new interaction is an additional contribution to the usual Zeeman splitting of energy levels. Their measurements led to the constraint $|{\cal A}|\lesssim 10^{-21}$.
  
  A 5th force magnetic field $\vec{\mathscr{B}}_{5,\oplus}'$ arising from the rotating Earth would produce a similar orientation-dependent contribution to the Zeeman splitting as in the Venema experiment.  However, in this case the interaction Hamiltonian would be of the form
  \begin{equation}
      H_{\rm int} = -g_{I}\mu_{N}\vec{I}\cdot\vec{B}  -\gamma_{5,I}\vec{I}\cdot\vec{\mathscr{B}}_{5,\oplus}',
      \label{hypermagnetic int}
  \end{equation}
  where $\gamma_{5,I}$ is the hypermagnetic gyromagnetic ratio of the nucleus. The contribution of the magnetic 5th force interaction in Eq.~(\ref{hypermagnetic int}) has the same form as the monopole-dipole interaction in Eq.~(\ref{monopole-dipole}), except $({\cal A}\epsilon/|\vec{I}|)\hat{r} \rightarrow -\gamma_{5,I}\vec{\mathscr{B}}_{5,\oplus}'$.

\section{Other Experiments, Systems, and Couplings}

\subsection{Satellite Tests of the Equivalence Principle}

Presently, the most stringent limit on violations of the WEP come from the MICROSCOPE satellite experiment \cite{Touboul 2022,Touboul 2022 CQG,Touboul 2017}, which has found the constraint on the differential acceleration parameter $\Delta\kappa_{ij} \lesssim 10^{-14}$ for accelerations relative to the Earth.  Since MICROSCOPE is in a Sun-synchronous, circular orbit at an altitude of 710~km above the Earth, it can be used to set significant constraints on longer ranged 5th forces \cite{Fayet 2018,Fayet 2019}, but not on a 5th force with range $\lambda_{5} \ll 710$~km using the Earth as the source. 

 MICROSCOPE is not fixed on the surface of the rotating Earth as in the case of terrestrial experiments, but is instead moving above the Earth at a relatively high orbital speed.  Therefore, the 5th force fields $\vec{\mathscr{E}}_{5,\oplus}$ and $\vec{\mathscr{B}}_{5,\oplus}$ acting on the satellite as a whole would be those relative to the inertial frame, given by Eqs.~(\ref{E_field}) and (\ref{B_field}).  If the hyperelectric field is suppressed, as in the case of the scalar-vector 5th force model of \ref{Scalar-Vector Appendix}, the dominant force acting on the satellite would be the hypermagnetic Lorentz force.  Since this force depends on $\vec{v}_{\rm sat}\times \vec{\mathscr{B}}_{5,\oplus}$, where $\vec{v}_{\rm sat}$ is the velocity of the satellite, its direction and magnitude will be more complicated than that of a simple 5th force that is radial in direction, as would be due to an electric 5th force, which is usually assumed.  However,  the test bodies in the satellite are themselves in a non-inertial frame.

  \subsection{Neutron Stars}
  
The focus of this paper has been on the effects of a magnetic 5th force arising from the rotation of the Earth.  However, other stellar-sized astrophysical bodies have considerably larger baryon numbers, and can have significantly larger rotation rates.  Probably the most extreme examples are neutron stars, which have been recognized as astrophysical laboratories which can be used to explore fundamental physics \cite{Pizzochero,Weber}. Typically, neutron stars have masses $M_{\rm NS}\gtrsim M_{\odot}$, radii $R_{\rm NS}\sim 10$~km, and can have rotational periods as small as $T_{\rm NS} \simeq 1.4$~ms \cite{Vidana}.  Then the ratio of the hypermagnetic field to the hyperelectric field would  be  approximately
    \begin{equation}
      \frac{|\vec{\mathscr{B}}_{5}|}{|\vec{\mathscr{E}}_{5}|} \sim  \frac{\omega_{\rm NS}R_{\rm NS}}{c}\sim 0.14,
    \end{equation} 
which is $\sim10^{6}$ greater than the value for the Earth.  Therefore, the hypermagnetic field  near a neutron star cannot be neglected relative to its hyperelectric field.  Significant 5th force tidal effects could also appear.  However, the enormous ordinary magnetic fields, $10^{4}$--$10^{8}$~T, that usually accompany a neutron star would need to be suppressed for the effects of a putative 5th force magnetic field to be observed.

An alternative approach to investigating the effects of a magnetic 5th force involving neutron stars is through its effects on their equation of state (EOS).  For the most part, previous work has focused on the impact of the Yukawa potential of the 5th force (i.e., the electric 5th force) on the EOS \cite{Song,Krivoruchenko,Wen}.  Multimessenger observations of neutron star mergers can be used to constrain models of the EOS \cite{Dietrich}, and could possibly also set limits on a magnetic 5th force.

\subsection{Accelerator Experiments}

As  described in Ref.~\citen{kaons}, part of the motivation leading to the suggestion of a 5th force were indications of possible anomalies in the decays of neutral kaons $K^0$, $\bar{K}^0$ (or $K_L,$ $K_S$). In principle, high energy kaons moving in the presence of a magnetic 5th force could experience an additional frame dependent contribution to their decay rates. To date, the implications of a magnetic 5th force on the $K^0$-$\bar{K}^0$ system have not been explored in detail either theoretically or experimentally \cite{kaons}. The same can be said for other high energy systems.

\subsection{Other Couplings}
    
The existence of another classical gauge field co-existing with the classical electromagnetic field then leads to a possibility of the coupling of these two fields.  An example of such a phenomenon can be found in the theory of  dark photons \cite{Fabbrichesi}.  Following this example, we can write the kinetic gauge field sector Lagrangian as
	\begin{equation}
	    {\cal L} = -\frac{1}{4}F^{\alpha\beta}F_{\alpha\beta} 
	     -\frac{1}{4}\mathscr{F}_{5}^{\alpha\beta}\mathscr{F}_{5,\alpha\beta}  -\frac{\varepsilon_{5}}{2}F^{\alpha\beta}\mathscr{F}_{5,\alpha\beta},
	\end{equation}
where $F^{\alpha\beta}$ is the usual electromagnetic field tensor, and $\varepsilon_{5}$ is the dimensionless coupling between the electromagnetic and 5th force fields.  The phenomenological consequences of this coupling, and the  coupling of the 5th force fields to other bosonic fields is beyond the scope of this paper.

Motivated by paper by Lee and Yang,\cite{Lee Yang} this paper has focused on a 5th force coupling to baryon number $B$.  Alternatively, total lepton number $L$ is another conserved quantity in the Standard Model, while lepton flavor is also conserved in situations that do not involve neutrino oscillations.  Therefore, one can also investigate a 5th force coupling to lepton number and its magnetic effects as we have done here.  

    \section{Discussion}
    
    In this paper, we investigated the phenomenological consequences of a magnetic 5th force coupling to baryon number, an idea which was first suggested, but not studied, by Lee and Yang in 1955 \cite{Lee Yang}.   After obtaining the general field  equations, which are analogous to Maxwell's equations with a massive photon, we calculated the hyperelectric and hypermagnetic fields for a rotating Earth.  We then transformed these fields to the rotating frame, which is appropriate for analyzing terrestrial experiments.  The magnetic 5th force acting on test bodies has a directional and position dependence that is significantly different from the corresponding electric 5th force, which has been the focus of various 5th force experiments.  If the electric 5th force is suppressed, as in the case of the scalar-vector model discussed in \ref{Scalar-Vector Appendix}, the magnetic 5th force could dominate.  This would then require re-evaluating the constraints on  the 5th force from previous experiments, and undertaking new experiments designed to detect the magnetic 5th force.

    One of the motivations for the present work has been to address what has become  known as the E\"{o}tv\"{o}s Paradox \cite{Fischbach PoS}: How could it be that so many modern experiments have failed to confirm the rather compelling evidence for a 5th force that has emerged from detailed analysis of the original EPF experiment? A possible response, suggested by the present work, is this:  5th force experiments are typically optimized to test a particular force model, which assumes a dependence on the coupling to matter, position-dependence,  direction-dependence, velocity-dependence, and source-dependence.  A good example is the MICROSCOPE satellite experiment  testing the weak equivalence principle (WEP), which is optimized to detect a violation of the WEP arising from the Earth \cite{Touboul 2022,Touboul 2022 CQG}.  It cannot be used to set limits on short-ranged 5th forces from the Earth, and its analysis protocol would need to be revised to constrain forces coming from other directions.  The magnetic 5th force discussed in this paper provides an example of a model where the force acting on a test body has a very different character than the electric 5th force, which heretofore has been assumed in most experiments.  This illustrates how even a horizontal force can arise on the surface of the idealized spherical Earth, when only a vertical force would normally be expected.  The dependence of local topography for a short-ranged magnetic 5th force, which was not discussed in this paper, would certainly be more complicated.  It is hoped that the example of the magnetic 5th force will motivate experimentalists to re-evaluate their approaches to constraining new macroscopic-ranged forces.  Not only would this further our understanding of physics beyond the Standard Model, and possibly lead to new methods for detecting dark matter, but it might also  provide a final resolution of the E\"{o}tv\"{o}s Paradox.

\section*{Acknowledgments}

We thank Quan Le Thien for useful discussions.

    \appendix

    \section{Potentials and Fields Produced by the Earth}
    \label{Potentials Fields Appendix}
    
    In this appendix, we provide the details for the calculations of the 5th force potentials and fields arising from the simple model of the rotating Earth, which assumes the baryon number density and current density are given by
\begin{equation}
    \rho_{5}(\vec{r}\,') = \left\{\begin{array}{ll}
        \rho_{5,\oplus} \equiv \displaystyle \frac{3B_{\oplus}}{4\pi R_{\oplus}^{3}}, & r'\leq R_{\oplus}, \\
        &\\
        0, & r' > R_{\oplus},
    \end{array}\right.
\end{equation}
and
\begin{equation}
    \vec{J}_{5}(\vec{r}\,') = \left\{\begin{array}{ll}
        \rho_{5,\oplus}\vec{\omega}_{\oplus} \times \vec{r}\,' = \rho_{5,\oplus}\omega_{\oplus}r'\sin\theta'\,\hat{\phi}' ,  & r'\leq R_{\oplus}, \\
        &\\
        0, & r' > R_{\oplus}.
    \end{array}\right.
\end{equation}
In this appendix, we choose units where $\hbar = c = 1.$

\subsection{Potentials}

\subsubsection{Scalar Potential}

To calculate the fields above the Earth's surface, where nearly all experiments are conducted,  we first calculate the 5th force scalar and vector potentials by combining Eqs.~(\ref{static Phi A}) and Eq.~(\ref{Ylm expansion}).  For $r' \leq R_{\oplus}$ and $r > R_{\oplus}$, Eqs.~(\ref{static Phi}) and (\ref{Ylm expansion}) give for the scalar potential,
\begin{equation}
    \Phi_{5}(\vec{r})  = f_{5}\mu_{5}\rho_{5,\oplus}
\sum_{l = 0}^{\infty} \sum_{m = -l}^{l} k_{l}(\mu_{5}r) Y_{l}^{m}(\theta,\phi)
\left[\int^{R_{\oplus}}_{0}dr'r'^{2} \,i_{l}(\mu_{5}r')\right]\left[\int d\Omega'\,Y_{l}^{m*}(\theta',\phi')\right].
\label{earth Phi}
\end{equation}
Using $1 = \sqrt{4\pi}\,Y^{0}_{0}(\theta,\phi)$,
\begin{equation}
\int d\Omega'\,Y_{l}^{m*}(\theta',\phi') = \sqrt{4\pi} \int d\Omega'\,Y_{l}^{m*}(\theta',\phi')
Y^{0}_{0}(\theta',\phi') = \sqrt{4\pi} \,\delta_{l,0}\,\delta_{m,0},
\end{equation}
Eq.~(\ref{earth Phi}) becomes
\begin{align}
\Phi_{5}(\vec{r})  &= \sqrt{4\pi}\, f_{5}\mu_{5}\rho_{5,\oplus}
\sum_{l = 0}^{\infty} \sum_{m = -l}^{l} k_{l}(\mu_{5}r) Y_{l}^{m}(\theta,\phi)
\left[\int^{R_{\oplus}}_{0}dr'r'^{2} \,i_{l}(\mu_{5}r')\right]\,\delta_{l,0}\,\delta_{m,0},
\nonumber \\
&=\sqrt{4\pi}\, f_{5}\mu_{5}\rho_{5,\oplus}
\, k_{0}(\mu_{5}r) Y_{0}^{0}(\theta,\phi)
\left[\int^{R_{\oplus}}_{0}dr'r'^{2} \,i_{0}(\mu_{5}r')\right],
\label{earth Phi 2}
\end{align}
where $i_l(x)$ and $k_l(x)$ are modified spherical Bessel functions of the first and second kind. In particular, we have 
\begin{subequations}
\label{k0 i0}
\begin{align}
k_{0}(x) &= \frac{e^{-x}}{x},\\
i_{0}(x) &= \frac{\sinh x}{x}.
\end{align}
\end{subequations}
The radial integral then becomes
\begin{align}
\Phi_{5}(\vec{r})  
&= f_{5}\mu_{5}\rho_{5,\oplus}
\,\left(\frac{e^{-\mu_{5}r}}{\mu_{5}r}\right) 
\int^{R_{\oplus}}_{0}dr'r'^{2} \,\left[\frac{\sinh(\mu_{5}r')}{\mu_{5}r'}\right],
\nonumber \\
&= f_{5}\mu_{5}\rho_{5,\oplus}
\,\left(\frac{e^{-\mu_{5}r}}{\mu_{5}r}\right) 
\left(\frac{\mu_{5}R_{\oplus} \cosh\mu_{5}R_{\oplus} - \sinh \mu_{5}R_{\oplus}}{\mu_{5}^{2}} \right),
\label{earth Phi 3}
\end{align}
which simplifies to
\begin{equation}
\Phi_{5}(\vec{r}) = \frac{f_{5}\rho_{5,\oplus}}{\mu_{5}^{2}}K_{E}(\mu_{5}R_{\oplus})
\left(\frac{e^{-\mu_{5}r}}{\mu_{5}r}\right),
\label{earth Phi Appendix}
\end{equation}
where 
\begin{equation}
    K_{E}(x) \equiv x\cosh x - \sinh x =\frac{x^{3}}{3} + {\cal O}(x^{5}).
\end{equation}
As a check, in the limit of a massless boson, we obtain the expected result
\begin{equation}
\lim_{\mu_{5}\rightarrow 0}\Phi_{5}(\vec{r}) = \frac{f_{5}\rho_{5,\oplus}R_{\oplus}^{3}}{3r} = \frac{f_{5}}{4\pi}\frac{B_{\oplus}}{r}.
\end{equation}

\subsubsection{Vector Potential}

The vector potential is obtained by combining Eqs.~(\ref{static A}) and (\ref{Ylm expansion}),
\begin{equation}
    \vec{\mathscr{A}}_{5}(\vec{r})  = f_{5}\mu_{5}\rho_{5,\oplus}\omega_{\oplus}
\sum_{l = 0}^{\infty} \sum_{m = -l}^{l} k_{l}(\mu_{5}r) Y_{l}^{m}(\theta,\phi)
\left[\int^{R_{\oplus}}_{0}dr'r'^{3} \,i_{l}(\mu_{5}r')\right]\left[\int d\Omega'\,Y_{l}^{m*}(\theta',\phi')\sin\theta'\,\hat{\phi}'\right].
\label{earth A}
\end{equation}
To carry out the angular integrals, we first write $\hat{\phi}'$ in terms of Cartesian unit vectors which are independent of position:
\begin{align}
\int d\Omega'\,Y_{l}^{m*}(\theta',\phi')\sin\theta'\,\hat{\phi}' 
&=\int d\Omega'\,Y_{l}^{m*}(\theta',\phi')\sin\theta'
\left(-\sin\phi'\,\hat{x} + \cos\phi'\,\hat{y}\right)
\nonumber \\
&=\int d\Omega'\,Y_{l}^{m*}(\theta',\phi')
\left[
-\sin\theta'\left(\frac{e^{i\phi'}-e^{-i\phi'}}{2i}\right)\hat{x}
+ \sin\theta'\left(\frac{e^{i\phi'}+e^{-i\phi'}}{2}\right)\hat{y}\right]
\nonumber \\
&=\int d\Omega'\,Y_{l}^{m*}(\theta',\phi')
\sqrt{\frac{8\pi}{3}}
\left\{\frac{1}{2i}\left[Y^{1}_{1}(\theta',\phi') + Y^{-1}_{1}(\theta',\phi')\right]\hat{x}
\right.
\nonumber \\
& \left.\mbox{} +\frac{1}{2}\left[-Y^{1}_{1}(\theta',\phi') + Y^{-1}_{1}(\theta',\phi') \right]\hat{y}\right\}
\nonumber \\
& \sqrt{\frac{8\pi}{3}}\delta_{l,1}\left[\frac{1}{2i}(\delta_{m,1} + \delta_{m,-1})\,\hat{x}
+ \frac{1}{2}(-\delta_{m,1} + \delta_{m,-1})\,\hat{y}
\right].
\label{A angular int}
\end{align}
Inserting Eq.~(\ref{A angular int}) back into Eq.~(\ref{earth A}) then gives
\begin{eqnarray}
   \vec{\mathscr{A}}_{5}(\vec{r})  &=& f_{5}\mu_{5}\rho_{5,\oplus}\omega_{\oplus}
 k_{1}(\mu_{5}r) 
\left[\int^{R_{\oplus}}_{0}dr'r'^{3} \,i_{1}(\mu_{5}r')\right]
\nonumber \\
&& \mbox{} \times
\sqrt{\frac{8\pi}{3}}\left\{
\frac{1}{2i}\left[Y_{1}^{1}(\theta,\phi) + Y_{1}^{-1}(\theta,\phi)\right]\hat{x}
+ \frac{1}{2}\left[-Y_{1}^{1}(\theta,\phi) + Y_{1}^{-1}(\theta,\phi)\right]\hat{y}
\right\} 
\nonumber \\
&=& f_{5}\mu_{5}\rho_{5,\oplus}\omega_{\oplus}
 k_{1}(\mu_{5}r) 
\left[\int^{R_{\oplus}}_{0}dr'r'^{3} \,i_{1}(\mu_{5}r')\right]
\left\{\sin\theta \left[
\left(\frac{-e^{i\phi} +e^{-i\phi}}{2i}\right)\hat{x}
+ \left(\frac{e^{i\phi} +e^{-i\phi}}{2}\right)\hat{y}
\right]\right\}, \nonumber \\
\end{eqnarray}
which simplifies to
\begin{equation}
\vec{\mathscr{A}}_{5}(\vec{r})  
= f_{5}\mu_{5}\rho_{5,\oplus}\omega_{\oplus}
 k_{1}(\mu_{5}r) 
\left[\int^{R_{\oplus}}_{0}dr'r'^{3} \,i_{1}(\mu_{5}r')\right]\sin\theta\,\hat{\phi},
\label{earth A 2}
\end{equation}
where
\begin{subequations}
\label{k1 i1}
\begin{align}
k_{1}(x) &= \frac{e^{-x}(x + 1)}{x^{2}},\\
i_{1}(x) &= \frac{x\cosh x- \sinh x}{x^{2}}.
\end{align}
\end{subequations}
The remaining radial integral is
\begin{align}
    \int^{R_{\oplus}}_{0}dr'r'^{3} \,i_{1}(\mu_{5}r') &=
    \frac{1}{\mu_{5}^{4}}
    \int^{\mu_{5}R_{\oplus}}_{0}dx
    \left(x^{2}\cosh x -x\sinh x\right),
\nonumber \\
&= \frac{1}{\mu_{5}^{4}}
\left[(3 + \mu_{5}^{2}R_{\oplus}^{2})\sinh\mu_{5}R_{\oplus} - 3\mu_{5}R_{\oplus}\cosh\mu_{5}R_{\oplus}\right].
\label{A radial int}
\end{align}
Combining Eqs.~(\ref{earth A 2})--(\ref{A radial int}) then gives the final result
\begin{equation}
\vec{\mathscr{A}}_{5}(\vec{r})  
= f_{5}\rho_{5,\oplus}\omega_{\oplus}
K_{B}(\mu_{5}R_{\oplus})
\frac{e^{-\mu_{5}r}(1 + \mu_{5}r)}{\mu_{5}^{5}r^{2}}
\sin\theta\,\hat{\phi},
\label{earth A Appendix}
\end{equation}
where
\begin{equation}
K_{B}(x) \equiv (3 + x^{2})\sinh x - 3x\cosh x =\frac{x^{5}}{15} + {\cal O}(x^{7}).
\end{equation}
In the limit of a massless boson, we find
\begin{equation}
\lim_{\mu_{5}\rightarrow 0}\vec{\mathscr{A}}_{5}(\vec{r})  
= \frac{f_{5}\rho_{5,\oplus}\omega_{\oplus}R_{\oplus}^{5}}{15r^{2}}
\sin\theta\,\hat{\phi}
= \frac{f_{5}}{4\pi}\frac{B_{\oplus}R_{\oplus}^2\omega_{\oplus}}{5r^{2}}\sin\theta\,\hat{\phi},
\end{equation}
which agrees with the analogous result of the vector potential produced by a rotating ball of electric charge \cite{Pierrus}.

\subsection{Hyperelectric and Hypermagentic Fields}

The static hyperelectric field $\vec{\mathscr{E}}_{5}(\vec{r})$ for $r > R_{\oplus}$ is obtained from the potential $\Phi_{5}(\vec{r})$, Eq.~(\ref{earth Phi Appendix}), using Eq.~(\ref{static E phi}):
\begin{align}
		\vec{\mathscr{E}}_5(\vec{r}) &= -\nabla\Phi_{5}(\vec{r})
		\nonumber \\
		&= -\frac{f_{5}\rho_{5,\oplus}}{\mu_{5}^{2}}
		K_{E}(\mu_{5}R_{\oplus})
\left(\frac{\partial}{\partial r}\frac{e^{-\mu_{5}r}}{\mu_{5}r}\right)\hat{r}
		\nonumber \\
		&= \frac{f_{5}\rho_{5,\oplus}}{\mu_{5}^{3}}K_{E}(\mu_{5}R_{\oplus})
\left(1 + \mu_{5}r\right)\left(\frac{e^{-\mu_{5}r}}{r^{2}}\right)\hat{r}.
	\end{align}
In the massless limit, we obtain the expected result:
\begin{equation}
    \lim_{m_{5}\rightarrow 0}\vec{\mathscr{E}}_5(\vec{r}) 
    = \frac{f_{5}\rho_{5,\oplus}R^{3}_{\oplus}}{3r^{2}}\,\hat{r} = \frac{f_{5}}{4\pi}\frac{B_{\oplus}}{r^{2}}\,\hat{r}.
\end{equation}

Similarly, the static hypermagnetic field $\vec{\mathscr{B}}_5(\vec{r})$ for $r > R_{\oplus}$ is obtained from the vector potential $\vec{A}_{5}(\vec{r})$, Eq.~(\ref{earth A Appendix}), using Eq.~(\ref{B curl A}).  Since  $\vec{\mathscr{A}}_{5}(\vec{r}) = \mathscr{A}_{5}(r,\theta)\,\hat{\phi}$, we find,
\begin{align}
	\vec{\mathscr{B}}_{5}(\vec{r}) &= \nabla \times \vec{\mathscr{A}}_{5}(\vec{r}) 
	\nonumber \\
	&= \frac{1}{r\sin\theta}\frac{\partial}{\partial\theta}\left[\mathscr{A}_{5}(r,\theta)\sin\theta\right]\hat{r}
	- \frac{1}{r}\frac{\partial}{\partial r}\left[r\mathscr{A}_{5}(r,\theta)\right]\hat{\theta}
	\nonumber \\
	& = \frac{f_{5}\rho_{5,\oplus}\omega_{\oplus}}{\mu_{5}^{3}} K_{B}(\mu_{5}R_{\oplus})\left(\left\{\frac{1}{r\sin\theta}
	\left[\frac{e^{-\mu_{5}r}(1 + \mu_{5}r)}{\mu_{5}^{2}r^{2}} \right]\frac{\partial}{\partial\theta}\sin^{2}\theta
	\right\}\hat{r}
	-\left\{\frac{\sin\theta}{r} \frac{\partial}{\partial r} \left[\frac{e^{-\mu_{5}r}(1 + \mu_{5}r)}{\mu_{5}^{2}r} \right]\right\}\hat{\theta}\right)
	\nonumber \\
	& = \frac{f_{5}\rho_{5,\oplus}\omega_{\oplus}}{\mu_{5}^{3}} K_{B}(\mu_{5}R_{\oplus})\left\{\left[
	\frac{2e^{-\mu_{5}r}(1 + \mu_{5}r)}{\mu_{5}^{2}r^{3}} \cos\theta
	\right]\hat{r}
	+\left[\frac{e^{-\mu_{5}r}(1 + \mu_{5}r+ \mu_{5}^{2}r^{2})}{\mu_{5}^{2}r^{3}} \sin\theta\right]\hat{\theta}\right\}
	\nonumber \\
	& = \frac{f_{5}\rho_{5,\oplus}\omega_{\oplus}}{\mu_{5}^{5}} K_{B}(\mu_{5}R_{\oplus})\frac{e^{-\mu_5r}}{r^{3}}\left[
	2(1 + \mu_{5}r)\cos\theta\,\hat{r}
	+ (1 + \mu_{5}r+ \mu_{5}^{2}r^{2}) \sin\theta\,\hat{\theta}\right].
\end{align}
In the massless limit, we find
\begin{align}
\lim_{\mu_{5}\rightarrow 0}\vec{\mathscr{B}}_{5}(\vec{r}) &=
\frac{f_{5}\rho_{5,\oplus}\omega_{\oplus}R_{\oplus}^{5}}{15r^{3}} \left[
	2\cos\theta\,\hat{r}
	+  \sin\theta\,\hat{\theta}\right]
\nonumber \\
	&= \frac{f_{5}}{4\pi}\frac{B_{\oplus}\omega_{\oplus}R_{\oplus}^{2}}{5r^{3}} \left[
	2\cos\theta\,\hat{r}
	+  \sin\theta\,\hat{\theta}\right],
\end{align}
which agrees with the corresponding result in electrodynamics \cite{Pierrus}.

 \section{Scalar-Vector 5th Force}
 \label{Scalar-Vector Appendix}
    
  As in electromagnetism, we have seen that a classical 5th force arising from a vector field could give rise to classical hyperelectric and hypermagnetic fields, $\vec{\mathscr{E}}_{5}(\vec{r})$ and $\vec{\mathscr{B}}_{5}(\vec{r})$.  Since most matter is electrically neutral, it is common to find situations where the magnetic field is easily observable.  However, for normal matter, the baryon number of a body never vanishes so the hyperelectric field of a 5th force dominates the hypermagnetic field for non-relativistic systems.  In this appendix we will consider a scalar-vector 5th force model in which the hyperelectric field is significantly suppressed so that the effects of the hypermagnetic field can produce significant contributions.

The possibility of new scalar and vector fields coexisting with gravity has been considered by many authors \cite{Scherk,Macrae,Goldman1,Goldman2,Nieto1,Nieto2,Ander,Hughes,Nieto3,Bars1,Bars2,Visser,Beverini,Ford,Fischbach exponential,Eckhard}.  The phenomenology of a scalar-vector 5th force is interesting because the scalar and vector potentials are attractive and repulsive, respectively, in the time-independent limit.  For the situation considered in this paper, we will consider the case that both  scalar and vector fields  couple to baryon number.  Therefore, the total  potential energy between two point particles \#1 and \#2 with baryon numbers $B_{1}$ and $B_{2}$, respectively, separated by a distance $r$ is (assuming $\hbar = c = 1)$
\begin{equation}
    V_{5}^{SV}(r) = -\frac{f_{5,S}^{2}}{4\pi}\frac{B_{1}B_{2}}{r}e^{-\mu_{5,S}r}
     +\frac{f_{5,V}^{2}}{4\pi}\frac{B_{1}B_{2}}{r}e^{-\mu_{5,V}r},
\end{equation}
where $f_{5,S}^{2}$ ($f_{5,V}^{2}$) and $\mu_{5,S}$ ($\mu_{5,V}$) are the scalar (vector) coupling constants and boson masses, respectively.  

Let us now consider a single nucleon interacting with a body with baryon density $\rho(\vec{r})$.  In the stationary approximation, the scalar and vector field potentials due to this body are obtained by integrating over the  body,
\begin{subequations}
\begin{align}
    \Phi_{5}^{S}(\vec{r})  &= -\frac{f_{5,S}}{4\pi} \int d^{3}r'\,\frac{e^{-\mu_{5,S}|\vec{r}-\vec{r}\,'|}}{|\vec{r}-\vec{r}\,'|}\rho(\vec{r}\,'),
\label{static S Phi}
    \\
    \Phi_{5}^{V}(\vec{r})  &= \frac{f_{5,V}}{4\pi} \int d^{3}r'\,\frac{e^{-\mu_{5,V}|\vec{r}-\vec{r}\,'|}}{|\vec{r}-\vec{r}\,'|}\rho(\vec{r}\,').
\label{static V Phi}
\end{align}
\end{subequations}
The total time-independent force on a single nucleon due to the body is then 
\begin{align}
    \vec{F}(\vec{r}) &= -f_{5,S}\vec{\nabla}\Phi_{5}^{S}(\vec{r}) - f_{5,V}\vec{\nabla}\Phi_{5}^{V}(\vec{r})
    \nonumber \\
    &= f_{5,S}\vec{\mathscr{E}}_{5,S}(\vec{r}) + f_{5,V}\vec{\mathscr{E}}_{5,V}(\vec{r}),
\end{align}
where $\vec{\mathscr{E}}_{5,S}(\vec{r})\equiv -\vec{\nabla}\Phi_{5}^{S}(\vec{r})$ is the effective scalar hyperelectric field, and $\vec{\mathscr{E}}_{5,V}(\vec{r})$ is the usual vector hyperelectric field due to the body.  

So far our treatment has been completely general. Now consider a theory where the coupling constants and boson masses have been fine-tuned (perhaps by some symmetry) such that the scalar and vector interactions have a common coupling constant ($f_{5,S} \simeq f_{5,V}$), and they either have approximately the same boson masses ($m_{5,S} \simeq m_{5,V}$), or both have sufficiently small masses to appear massless at the scale of experiments or astronomical observations.  Since the stationary scalar and vector hyperelectric fields due to a body have opposite signs, they will cancel. Hence, the hyperelectric fields will be strongly suppressed relative to the vector hypermagnetic field, for which there is no scalar counterpart.  Therefore, in such a theory, the magnetic 5th force will become the dominant force on a nucleon.

\end{document}